\documentclass[12pt]{article}
\usepackage[utf8]{inputenc}
\usepackage{graphicx} 
\usepackage{amsmath}
\usepackage[labelfont=bf]{caption}
\usepackage{setspace}
\usepackage[letterpaper, portrait, margin=1in]{geometry}
\usepackage[nottoc,numbib]{tocbibind}

\usepackage{amssymb}
\usepackage[labelfont=bf]{caption}
\usepackage{subfig}
\usepackage{setspace}
\usepackage{titlesec}
\usepackage{wrapfig}
\usepackage{float}
\usepackage{cite}
\usepackage{url}

\newcommand{\beginsupplement}{
    \setcounter{section}{0}
    \renewcommand{\thesection}{S\arabic{section}}
    \setcounter{equation}{0}
    \renewcommand{\theequation}{S\arabic{equation}}
    \setcounter{figure}{0}
    \renewcommand{\thefigure}{S\arabic{figure}}
    \setcounter{table}{0}
    \renewcommand{\thetable}{S\arabic{table}}
    \newcounter{offset}
  \setcounter{offset}{\value{figure}}
  \renewcommand{\thefigure}{S\the\numexpr\value{figure}-\value{offset}\relax}
}

\usepackage{hyperref}

\titleformat{\section}[block]
  {\normalfont\large\bfseries}{\thesection}{1em}{\large}
  
\title{Preserved reptile scales retain microscopic features, revealing a new instance of convergent evolution}

\author{Calvin A. Riiska$^{1}$, Gordon W. Schuett$^{2,3}$, \\ Joseph R. Mendelson III$^{4,5}$,  Jennifer M. Rieser$^{1}$\\
\small $^{1}$Emory University Department of Physics, Atlanta, GA\\
\small$^{2}$Department of Biology | Neuroscience Institute, Georgia State University, Atlanta, GA\\
\small$^{3}$Chiricahua Desert Museum, Rodeo, NM\\
\small$^{4}$Department of Research, Zoo Atlanta, Atlanta, GA\\
\small$^{5}$School of Biological Sciences, Georgia Institute of Technology, Atlanta, GA\\}
\date{\today }

\begin{document}

\maketitle

\section*{Abstract}
Small-scale structures on biological surfaces can profoundly impact how animals move, appear, and interact with their environments. Such textures may be especially important for limbless reptiles, such as snakes and legless lizards, because their skin serves as the primary interface with the world around them. Here, we examine ventral microstructures of several limbless reptiles, which are hypothesized to be highly specialized to aid locomotion via frictional interactions. Inspired by prior studies that investigated potential links between microtextures, phylogeny, habitat, and locomotion---but that were limited by their reliance on shed skins---we characterized the structures present on preserved museum specimens and found that they are quantitatively similar to those found on shed skins. Using this result, we confirmed a previously hypothesized---but untested due to the lack of shed skins---third independent evolution of sidewinding-specific isotropic microtexture. Specifically, we examined a museum-preserved \textit{Bitis peringueyi} specimen and identified a new instance of convergent evolution in sidewinding viper microstructures: the loss of micro-spikes (present on many snake species) and the appearance of micro-pits with a characteristic spacing. Our results reveal that museum-preserved specimens retain intact microtextures, greatly expanding the availability of samples for evolutionary studies.

\section{Introduction}
The diverse textures present on animal skins or exoskeletons often play crucial roles in mediating interactions with the environment. Remarkably, microtextures with features on the micron or nanometer scale, often requiring the use of sub-diffraction-limit imaging techniques to fully characterize, can influence environmental interactions at macroscopic  scales by modulating interactions with light, sound, fluids, and other biological matter~\cite{riiska_physics_2024}. For example, reflective microstructures on birds and butterflies can enhance coloration for mating displays~\cite{mccoy_microstructures_2021}, reduce visibility for predator avoidance~\cite{vukusic_photonic_2003,vieirasilva_relevance_2024}, and aid in thermoregulation~\cite{krishna_infrared_2020}. Sound absorbing micro-scales on moth wings provide acoustic camouflage from predatory bats~\cite{shen_biomechanics_2018,neil_moth_2020,neil_moth_2022}. Hydrophobic textures help shed water from insect wings~\cite{sharma_anti-wetting_2025,sun_structure_2011}, and when combined with hydrophilic structures, can enable water harvesting in both desert beetles~\cite{parker_water_2001} and rattlesnakes \cite{phadnis2019role} as well as adhesion to air-water interfaces~\cite{suzuki_hydrophobic-hydrophilic_2021}. Micron-sized denticles on shark skins help to streamline flows around their bodies~\cite{popp_denticle_2020,lauder_structure_2016}, enhance thrust~\cite{savino_thrust_2024}, and protect delicate tissues like gill flaps~\cite{gabler-smith_dermal_2021}. Microstructures can also interlock (as in bird feathers~\cite{matloff_how_2020}, enhancing lift during flight) or prevent adhesion (as in carnivorous pitcher plants~\cite{labonte_disentangling_2021}, aiding in prey capture by destabilizing insect footholds). 

Microstructures can also directly enhance locomotion through specialized interactions with other surfaces. For example, setae on gecko feet utilize van der Waals forces to allow these animals to adhere to both smooth and rough surfaces~\cite{autumn_mechanisms_2002,huber_influence_2007,tian_adhesion_2006}. Similar structures are found on the feet of flies and beetles and, when combined with a sticky fluid secretion, these insects can cling to a variety of surfaces as well~\cite{beutel_ultrastructure_2001,dirks_fluid-based_2011,gorb_design_1998,schroeder_its_2018}. Other insects such as ants, stick insects, and katydids have smooth pads with internal fibers that distort around surface features to increase contact-area-dependent forces that are further enhanced by the presence of a fluid secretion~\cite{schroeder_its_2018,beutel_ultrastructure_2001,dirks_fluid-based_2011,labonte_functionally_2013,grohmann_two_2015}.

For limbless animals moving across terrestrial substrates, the ventral surface serves as the primary contact with the environment during locomotion. Consequently, it has been hypothesized that specialized ventral microstructures likely evolved to enhance frictional interactions. Many snake species have rearward-pointing micro-spikes that are thought to aid movement through the creation of anisotropic friction---providing low friction for forward sliding, high friction to prevent backward sliding~\cite{baum_anisotropic_2014}, and high lateral friction to reduce sideways slip and enhance forward locomotion~\cite{benz2012anisotropic, hu2009mechanics}. Intriguingly, these structures also appear to be linked to habitat; qualitative observations suggest that arboreal snakes have longer spikes than terrestrial species, potentially enhancing friction forces that are crucial for traversing trees, while aquatic and marine species tend to have shorter spikes, perhaps due to the decreased importance of interactions with solid substrates~\cite{schmidt_snake_2012}. Fossorial squamates are unique in that the dorsal surfaces also interact with the substrate and may be smoother than the dorsal scales of non-fossorial species~\cite{martinez_quantifying_2021}, possibly to reduce friction during burrowing and subterranean locomotion~\cite{sharpe_locomotor_2015}. Notably, for some desert-dwelling sidewinding vipers, micro-spikes are either greatly reduced or completely absent. Instead, they possess an isotropic array of pits, found to have evolved independently at least twice in distantly related species of vipers ~\cite{rieser_functional_2021}. Computational modeling revealed that sidewinding, a periodic pattern of movement in which a planar wave of body curvature is coupled to a vertical wave of body lifting (where lifted segments are moved forward and non-lifted segments are pressed against the substrate) that produces an overall sideways motion~\cite{gans_kinematic_1992,jayne_kinematics_1986}, is enhanced by isotropic friction, which these isotropic structures are hypothesized to produce~\cite{rieser_functional_2021}.

While these results strongly suggest an important interplay between microstructures in squamate reptiles (as well as in fishes, e.g.,~\cite{wainwright_imaging_2017}, though not explored here), movement strategies, habitats, and phylogeny, establishing definitive links has been constrained by limited sample availability; previous studies have largely relied on shed skins~\cite{rieser_functional_2021,hazel_nanoscale_1999,reza_nanopores_2025,spinner_non-contaminating_2014,arrigo_phylogenetic_2019}, which can be difficult to obtain for rare or elusive squamate species. 

\begin{figure}
    \centering
    \includegraphics[width=\textwidth]{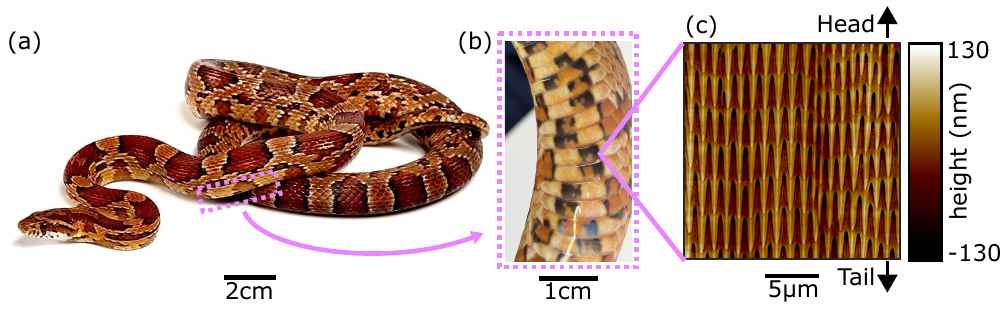}
    \caption{\textbf{The ventral scales of snakes possess microscopic textures.} \textbf{(a)} A cornsnake (\textit{Pantherophis guttatus}). \textbf{(b)} Zoomed-in image of several ventral scales, which serve as the primary animal-substrate interface during locomotion. \textbf{(c)} Microstructures within a 20~$\mu$m~$\times$~20~$\mu$m region of a single ventral scale from the shed skin of a cornsnake, imaged using atomic force microscopy. Many snake species have similar textures, composed of repeating rows of head-to-tail oriented spike-like features, that are thought to enhance frictional interactions during movement.}
    \label{fig:intro}
\end{figure}

In contrast, natural history museums house extensive collections of formalin-fixed and ethanol-preserved specimens of all forms on non-avian reptiles, samples of which are made available to researchers upon request for myriad avenues of of biological research~\cite{dodd_preserving_2016,jacobson2021overview},
including biometric (e.g., limb sizes, bone shapes, and allometry)~\cite{sokal1995biometry} and histological (e.g. cell arrangements and tissue strength)~\cite{musumeci2014past} measurements, information on basic natural history and life habits~\cite{reiserer2018seed}, and DNA extraction~\cite{zacho_uncovering_2021} for genome sequencing to understand the origin of traits~\cite{pareek_sequencing_2011} or to recover phylogenetic relationships~\cite{alencar_diversification_2016}. While the preservation process is known to have some effects on reptile specimens, e.g., shortened snout-vent lengths, lower body masses~\cite{vervust_effect_2009}, and reduced pigmentation~\cite{jacobson2021overview}, the effects of preservation on fine-scale surface microstructures are not known. 
Some previous studies have included measurements of microstructures on preserved skin samples~\cite{berthe_surface_2009,schmidt_snake_2012,martinez_quantifying_2021,irish_scanning_1988,baeckens_ontogenetic_2019}, however, whether the geometry of preserved structures differs from those in sheds was not explored or discussed. Given the stiffness of snake skin, which is composed of keratin~\cite{ripamonti_keratin-lipid_2009,toni_alpha-_2007,mckittrick_structure_2012}, we hypothesize that these microstructures may remain intact and comparable to other sample types through the processes of specimen fixation and preservation and thus be useful for analyses using AFM. If true, this would open extensive museum collections to quantitative analysis of keratin microstructures across diverse species~\cite{suarez_value_2004}, thus alleviating the limitation of sampling based on availability of shed skins and allowing comparisons with studies that have used other sample origins.

In this study, we used atomic force microscopy (AFM) to demonstrate that formalin-fixed, ethanol-preserved museum specimens retain microstructures comparable to those present on shed skins. Unlike the more widely used scanning electron microscopy (SEM)~\cite{russell_sem_nodate,klein_ultrastructure_2014,arrigo_phylogenetic_2019,schmidt_snake_2012}, AFM provides nanometer-resolution topographical measurements that enable precise quantification of in-plane geometry, vertical height profiles, and spatial anisotropy. SEM also requires sample surfaces to be sputtered with a metal coating such as silver or gold~\cite{reza_nanopores_2025,spinner_non-contaminating_2014,arrigo_phylogenetic_2019}, and it is unknown how this affects sample topology. Gel-based stereo profilometry and white-light 3D scans provide accurate reconstructions of raw surface topology in a variety of sample types, including live animals~\cite{wainwright_imaging_2017} (although their use is challenging for some species of squamates (e.g. vipers) that are difficult or dangerous to handle) and samples from taxa that do not shed skin layers (e.g., fishes)~\cite{baeckens_ontogenetic_2019,wainwright_imaging_2017,martinez_quantifying_2021}. Despite the versatility of these approaches, a significant limitation is their resolution---they are unable to capture the sub-diffraction limit microstructures of interest in our study. Therefore, we use AFM as it provides us with the highest resolution 3D surface quantification possible without altering the surface in any way.

Using the AFM-based approach described below, we quantitatively assess both the height distributions and dominant spacings of features within skin microstructures. We first show that textures on shed skins of cornsnakes, \textit{Pantherophis guttatus}, are similar within an individual (i.e., from distinct regions within a single scale), across individuals (i.e., from sheds collected from two different snakes), as well as for two preserved museum specimens. We then demonstrate that this structural similarity holds for multiple snake species as well as for a legless lizard (\textit{Ophisaurus attenuatus}), and conclude that microstructures are retained over decades throughout the preservation and storage processes. Finally, we leverage these results by using preserved samples to test the previously framed hypothesis~\cite{rieser_functional_2021}, that a new, fourth instance of the pitted microstructure associated with sidewinding may occur in Peringuey's adder \textit{Bitis peringueyi}, a species known to use sidewinding locomotion \cite{tingle_facultatively_2020}. If confirmed, this finding would reveal a third independent evolutionary origin of sidewinding behavior in snakes accompanied by a convergent morphology involving the loss of micro-spikes in favor of an isotropic pitted microstructure. Further, our test provides a concrete example of how museum collections can expand sample availability for studies using AFM to enable hypothesis testing and broader comparative studies of the qualities and evolution of specialized ventral microstructures, ultimately generating unique insights that could establish new links between behavior, habitat, and phylogeny.

\section{Results}

\begin{figure}
    \centering
    \includegraphics[width=\textwidth]{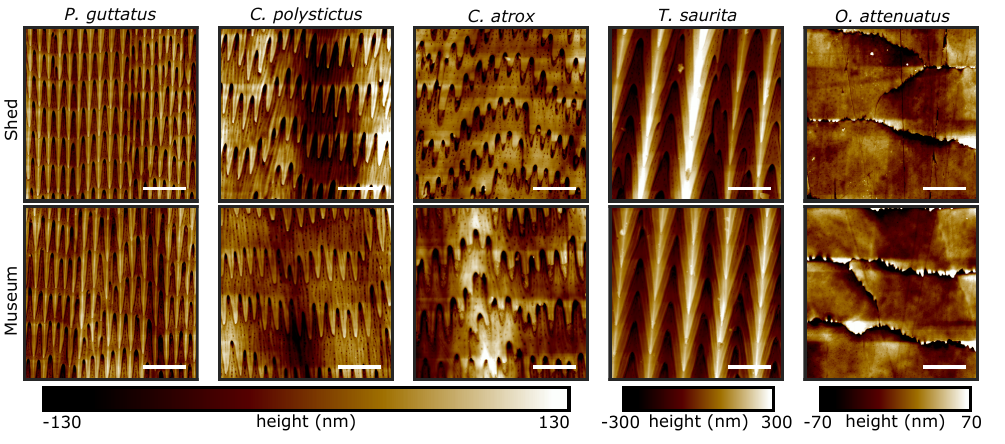}
    \caption{\textbf{Microstructure topographies for shed skins and preserved samples.} AFM scans of 20 $\mu$m x 20 $\mu$m regions on shed skins (top row) and museum-preserved samples (bottom row). Structures appear similar for four snake species (\textit{P. guttatus}, \textit{C. polystictus}, \textit{C. atrox}, and \textit{T. saurita}, left four columns) and a glass lizard (\textit{O. attenuatus}, right column). Microstructures are retained even for older samples---the \textit{C. polystictus} sample was preserved in 1939. Scale bars: $5~\mu$m, and the top of each image is toward the head of the animal.}
    \label{fig:visual}
\end{figure}

We used AFM to image $20$~$\mu$m~$\times$ $20$~$\mu$m sections on snake ventral scales (see Methods). These images revealed that the surface features and detailed morphologies of microstructures on preserved specimens closely resembled those on shed skin samples (Figure \ref{fig:visual}). We examined museum samples and shed samples from four snake species---the colubrids (\textit{Pantherophis guttatus}  (cornsnake) and \textit{Thamnophis saurita} (ribbonsnake)), and the viperids (\textit{Crotalus polystictus} (Mexican lance-headed rattlesnake) and \textit{Crotalus atrox} (western diamond-backed rattlesnake)). In each case, the ventral microstructures consisted of periodic rows of raised, rearward-pointing, overlapping spikes, typically $1-5$~$\mu$m in length. AFM enabled precise height measurements at the nanometer scale, revealing that spikes in \textit{P. guttatus}, for example, were raised about 100~nm above the surface. Secondary features such as ridges and small pits are observed between spikes or between rows of spikes. These microstructures are remarkably well preserved, even in decades-old specimens; our oldest sample, from \textit{C. polystictus}, was collected and preserved in 1939. 

We also observed similar keratinous microstructures on another squamate reptile. In ventral scales from both shed skin and museum samples of the legless slender glass lizard \textit{Ophisaurus attenuatus}, the shapes of cell borders and sub-micron-scale pits appear similar in both samples. These observations suggest that a wide range of keratin surface structures can be studied using preserved specimens across a broad taxonomic diversity of reptiles, greatly expanding sample availability for comparative and evolutionary studies.

\subsection{Quantitative similarity within and across shed skins}

\begin{figure}
    \centering
    \includegraphics[width=\textwidth]{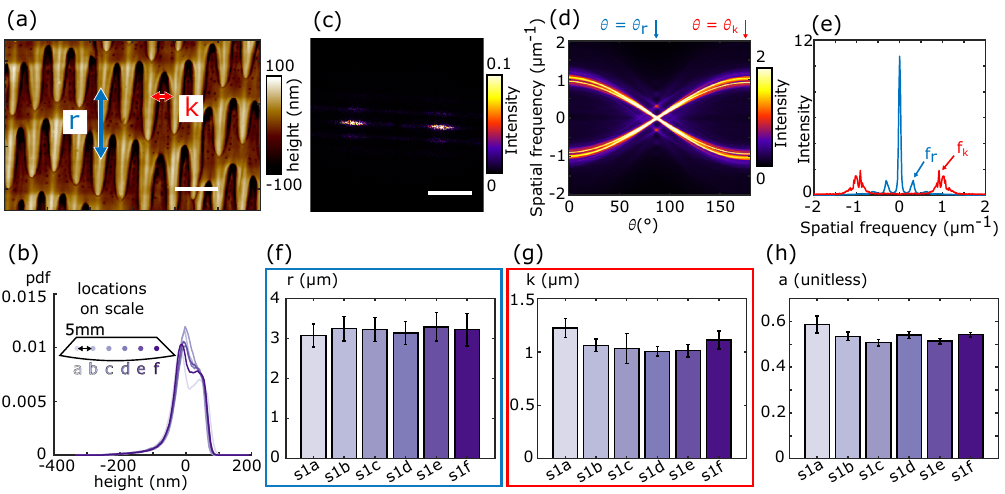}
    \caption{\textbf{Quantification of microscopic features reveals structural similarity across a single scale.} \textbf{(a)} A zoomed-in view of microstructures on a \textit{P. guttatus} shed skin (s1). Two quantities, $r$ and $k$ (the average row and spike spacing, respectively), are used to characterize and compare samples. Scale bar: $2$~$\mu$m. \textbf{(b)} A third quantity, the relative height variations within a sample, characterized through distributions of mean-subtracted feature heights, provides an additional comparative measurement. Distributions of relative height variations from evenly spaced sites within a single scale (as illustrated by the inset) indicate similarities across sites. \textbf{(c)} Power spectrum of zoomed-out version of (a) (see s1c image in Figure~\ref{fig:ims}), with dominant and secondary peaks associated with in-plane structural periodicities along and across spikes. Scale bar: $1$~$\mu$m$^{-1}$. \textbf{(d)} Radon transform of (c). The angular direction, $\theta_r$, where the two primary peaks collapse is identified as the angle with maximal intensity. \textbf{(e)}  Slices of the Radon transform used to calculate $r$ and $k$. Secondary peaks in the dominant direction slice ($\theta_r$, blue) correspond to the spatial frequency of rows of spikes. Peaks in the slice where the primary peaks collapse furthest apart ($\theta_k$, red) correspond to the lateral spatial frequency of the spikes. The \textbf{(f)} row spacing $r$, \textbf{(g)} spike spacing $k$, and \textbf{(h)} anisotropy indices $a$, are all similar across the six sites in the sample. In \textbf{(f)} and \textbf{(g)}, error bars represent the sample variability estimated using the standard deviation of the Gaussian fit to the associated slices at each site. In \textbf{(h)}, error bars represent the standard deviation of the anisotropy values in each of the four image quadrants.}
    \label{fig:spacing}
\end{figure}

To determine the extent to which ventral microstructures are similar within and across samples, as well as across sample types, we developed a robust quantification method inspired by previous work~\cite{rieser_functional_2021}. All analyses began with AFM-acquired height maps of $60~\mu$m~$\times~60~\mu$m regions. We started by quantifying features on a shed skin sample (s1) from \textit{P. guttatus}, chosen because of its common use in locomotion studies (e.g.,~\cite{jayne_why_2015,astley_arboreal_2009,jurestovsky_generation_2021,mansfield_arboreal_2011,marvi_friction_2012})and characterizing its microstructure may elucidate links to movement. To quantify the geometric arrangement of microstructural features, we first Gaussian filtered each image to remove depressions and bulges occurring over tens of microns caused by irregularities in the sample (see Figures \ref{fig:OGims}, \ref{fig:ims}, \ref{fig:atroxImsOG}, and \ref{fig:atroxIms}). These variations occur over much longer length scales (at least tens of microns) than the spiked structures we are interested in characterizing; removing these larger-scale variations allows for direct comparisons of the heights and spacings of the spiked features of interest. The resulting mean-subtracted image displays height variations associated with in-plane microstructures that can be compared across sample types. To assess within-sample variability, we imaged six distinct regions (denoted a--f, spaced about $5$~mm apart) along a single  ventral scale from a shed skin sample of \textit{P. guttatus}(Figure~\ref{fig:spacing}a). To compare height distributions across images, we used the Jensen-Shannon (JS) divergence, which is bounded between zero and one and measures the distinguishability of two distributions. A JS divergence value of zero indicates that two distributions are identical while a value of one indicates that the distributions are perfectly distinguishable. We found that the height distributions across these sites were similar (Figure \ref{fig:spacing}b), with low JS divergence values ($\leq0.04$) between each pair of distributions (see Table \ref{table:heightJS}).

For in-plane structure, we expanded previous analyses~\cite{rieser_functional_2021} to quantify two key spatial metrics: the spacing between the rows of spikes, $r$, and the lateral spacing between adjacent spikes, $k$ (see labels in Figure~\ref{fig:spacing}a). Starting with our Gaussian filtered, mean-subtracted height map (Figure~\ref{fig:ims}), we first computed the $2$D power spectrum (Figure~\ref{fig:spacing}c). The frequencies associated with the two brightest peaks in the power spectrum correspond to high lateral periodicity (the spike spacing, $k$), while secondary peaks above and below the primary peaks reflect vertical periodicity (the row spacing, $r$). The broadness of the primary and secondary peaks as well as the slight rotational offset---caused by imperfections in sample mounting---made extracting dominant peak locations and orientations from these images challenging; therefore, we used a Radon transform, which projects the intensity of a $2$D image onto lines of varying orientation that pass through the center of the image, to identify the directions associated with dominant peaks (see Section \ref{Analysis: Quantifying the structures} for more details).

The Radon transform of the power spectrum (Figure~\ref{fig:spacing}d) is brightest for the orientation, $\theta_r$, that projects the primary intensity peaks in the power spectrum onto the origin. In the slice at $\theta_r$, the secondary peaks above and below the primary power spectrum peaks collapse, creating smaller off-center peaks that identify the spatial frequency associated with the vertical spacing, $r$, between rows of spikes. The orientation, $\theta_k$, with the largest spacing between primary peaks in the Radon transform corresponds to the spatial frequency associated with the lateral spacing, $k$, between adjacent spikes.

Vertical slices through the Radon transform, taken at $\theta_r$ and $\theta_k$, are shown in Figure~\ref{fig:spacing}e. Row and spike spatial frequencies and variability were estimated by locally fitting Gaussians to relevant peaks (smaller off-center peaks for rows and off-center peaks for spikes; see Methods Section~\ref{Methods}\ref{Analysis: Quantifying the structures} and Figure~\ref{fig:calculation} for full details). The peak of each Gaussian fit (i.e., the mean value) represents the average spacing across the sample, and the width of the Gaussian (i.e., the standard deviation) represents the variability in these quantities across the sample. We also report uncertainties associated with the fit quality for each Gaussian which provides an estimate for how accurately we can measure the sample mean and variability. We note that the fit uncertainties remain small compared to sample variability for all samples studied (Table~\ref{table:allData}). At site s1c, the row spacing is $r=3.2 \pm 0.3\ \mu$m, and the spike spacing is $k=1.0 \pm 0.1\ \mu$m. This approach allows for systematic comparisons across sites, and across six sites within one sample, variation across sites fell within the sample variability (standard deviations) within each site. This is reflected by the bar graphs in Figure~\ref{fig:spacing}f and g---for each site, the height of each bar indicates the mean spacing value (given by peak of the Gaussian fit), and the error bar represents the sample variability (given by the standard deviation of the Gaussian fit). We conclude from these plots that there is no systematic variation across sites, and that the variation of the mean values across sites is comparable to the variability within a site. 

To quantify the directional variability of the microstructure, we measured the anisotropy index, $a$, a quantity introduced in a previous study~\cite{rieser_functional_2021}. The anisotropy index of a region is defined as the JS divergence between vertical slices of the Radon transform (of the power spectrum) separated by $90^\circ$. An anisotropy index close to zero indicates a uniform structure with little direction dependence while larger values indicate higher directionality. In our calculation, we divided each image into four quadrants and calculated the average and standard deviation of the anisotropy index over all quadrants. The anisotropy index at site s1c was $a=0.51\pm0.01$ indicating a large degree of directionality. While anisotropy standard deviations do not always overlap, the anisotropy values are comparable across sites, providing a baseline for structural variability in like samples.

A second shed skin sample (s2) from \textit{P. guttatus} showed that the height distribution and spacing of structures do not vary between individuals (Figure~\ref{fig:sample}c-e). This sample had a row spacing of $r=3.3 \pm 0.3\ \mu$m and a spike spacing of $k=1.1\pm 0.1\ \mu$m, nearly identical to s1c. Structures from different individuals were also similarly anisotropic---s2 had a calculated anisotropy of $0.62\pm0.02$---confirming that microstructural features are consistent between individuals of the same species (Figure \ref{fig:sample}f).

\subsection{Microstructures are retained on preserved samples}

\begin{figure}
    \centering
    \includegraphics[width=\textwidth]{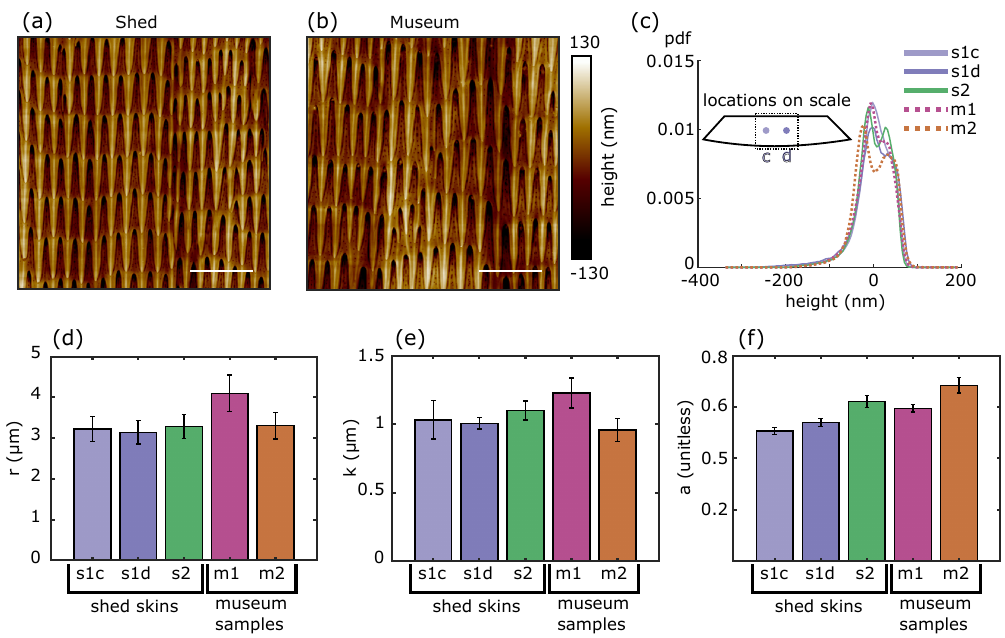}
    \caption{\textbf{Quantification of structural similarities across samples and sample types.}  AFM images of \textit{P. guttatus} microstructures of \textbf{(a)} a shed skin and \textbf{(b)} a preserved museum sample are visually similar. Scale bars: $5~\mu m$. \textbf{(c)} Distributions of height variations from two different individuals each for shed skins (s1 and s2) and museum samples (m1 and m2). The \textbf{(d)} row spacing and \textbf{(e)} spike spacing are quantitatively similar across sample types. \textbf{(f)} The anisotropy indices are slightly more variable but exhibit no systematic trend.}
    \label{fig:sample}
\end{figure}

To assess similarities and differences across sample types, we compared microstructures from two shed skins and two preserved samples of \textit{P. guttatus} (Figure~\ref{fig:sample}a,b). Distributions of feature heights (Figure~\ref{fig:sample}c) were consistent across all samples, with JS divergences~$\leq0.04$, comparable to within-sample variation. Use of spatial frequency analyses described above showed that preserved specimens retained row- and spike-spacing values similar to those from shed skins (Figure~\ref{fig:sample}d,e). While one museum specimen (m1) showed slightly higher average spacings ($r = 4.1 \pm 0.5\ \mu$m; $k = 1.2 \pm 0.2\ \mu$m), these values fell within the range of natural variation. The second museum specimen (m2) closely matched the shed skin measurements ($r = 3.3 \pm 0.4\ \mu$m; $k = 0.9 \pm 0.1\ \mu$m). Given that all measurements of $r$ and $k$ are within our calculated variability of each other, we find that these values are retained through the preservation process and remain comparable to shed skin samples. While the $a$ values of shed skins and museum samples do not fall within a sample standard deviation of each other, they do not vary more than our baseline variability established by comparing six sites on the same sample. Differences in $r$, $k$, and $a$ values can be attributed to natural variation in the samples.
It is worth noting that we also examined whether body size might influence structural spacing. Using ventral scale width as a proxy, we ranked specimens from largest to smallest: s1 ($31.7$~mm), m1 ($25.0$~mm), s2 ($17.9$~mm), and m2 ($12.6$~mm). No trend emerged linking individual size to microstructural spacing.

\begin{figure}
    \centering
    \includegraphics[width=\textwidth]{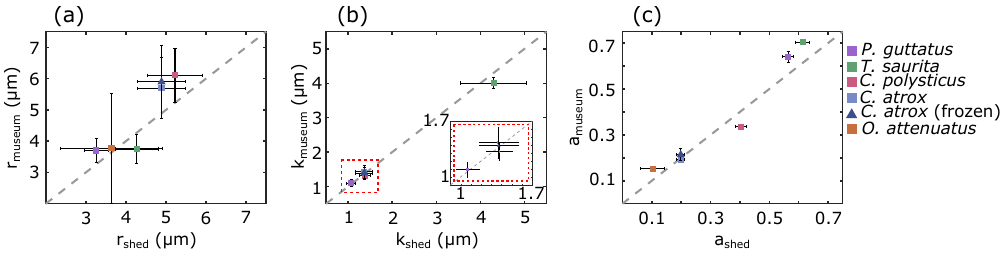}
    \caption{\textbf{Quantitative comparison of structural measurements across sample types for several species.} \textbf{(a)} The  preserved vs shed row spacings for four snake species and one lizard species all lie close to the dashed gray $y=x$ line. \textbf{(b)} The preserved vs shed spike spacings also lie close to $y=x$, indicating similarity of different sample types and multiple species. \textbf{(c)} The anisotropy indices are also similar across sample types (including a frozen sample for \textit{C. atrox}, blue triangle).  Note that \textit{O. attenuatus} only has a dominant structural frequency in the head-to-tail direction.}
    \label{fig:species}
\end{figure}

Extending our analysis beyond \textit{P. guttatus}, we applied the same quantification technique to preserved samples from three additional snake species---\textit{C. polystictus}, \textit{C. atrox}, and \textit{T. saurita}---and a glass lizard, \textit{O. attenuatus}. In all snake species, preserved samples had row and spike spacings and anisotropy indices consistent with those of shed skins (Figure~\ref{fig:species}). In \textit{O. attenuatus}, which only showed a dominant periodicity in the head-to-tail direction, both preserved and shed samples exhibited similar cell row dimensions and anisotropy indices, supporting the conclusion that keratin-based microtextures are preserved across diverse reptilian taxa. Table \ref{table:allData} summarizes measurements for each sample studied, including the sample average row spacing $r$ with associated sample variability $\sigma_r$ and fit uncertainty $\sigma_{rfit}$, the sample average spike spacing $k$ with associated sample variability $\sigma_k$ and fit uncertainty $\sigma_{kfit}$, and the anisotropy index with associated sample variability $\sigma_a$.

\subsection{A third independent origin of pitted microstructure in sidewinding snakes}

Statistical ancestral state reconstruction of specialized sidewinding---as opposed to idiosyncratic instances of sometimes poorly manifested sidewinding in unusual or non-natural situations---has independently evolved five times in vipers~\cite{tingle_facultatively_2020}. These include once in pitvipers (subfamily Crotalinae), in the American sidewinder rattlesnake (\textit{Crotalus cerastes}), and four times in the Old-World true vipers (subfamily Viperinae) including \textit{Echis carinatus}, the genus \textit{Cerastes} (three species), the common ancestor of the genera \textit{Pseudocerastes} (three species) and \textit{Eristocophis} (one species), and the common ancestor of a clade comprising the species \textit{Bitis caudalis}, \textit{Bitis peringueyi}, and \textit{Bitis schneideri} (the latter species is not considered a sidewinding specialist, thus it is a case of a potential evolutionary reversal per Tingle, 2020).   
Using our result that microstructures are preserved on museum specimens, we analyzed a museum-preserved sample of \textit{Bitis peringueyi}, a sidewinding true viper native to the Namib Desert of southern Africa~\cite{broadley1983fitzsimons}. We found that its ventral microstructure closely resembles those of other sidewinding viper species. Previous work~\cite{rieser_functional_2021} with the pitviper (Crotalinae)\textit{ Crotalus cerastes}, and the true vipers (Viperinae) \textit{Cerastes cerastes} and \textit{Cerastes vipera} identified the distinctive pitted microstructure and with the reduced or missing rearward-facing micro-spikes that are associated with specialized sidewinding. However, that study was limited to shed skin samples, and no shed skins were available for \textit{Bitis peringueyi}. By leveraging our results from this study and using a museum-preserved specimen, we were able to examine the relatively unrelated \textit{B. peringueyi}, thus expand the evolutionary context of sidewinding microstructures in four species of vipers. 

\begin{figure}
    \includegraphics[width=\textwidth]{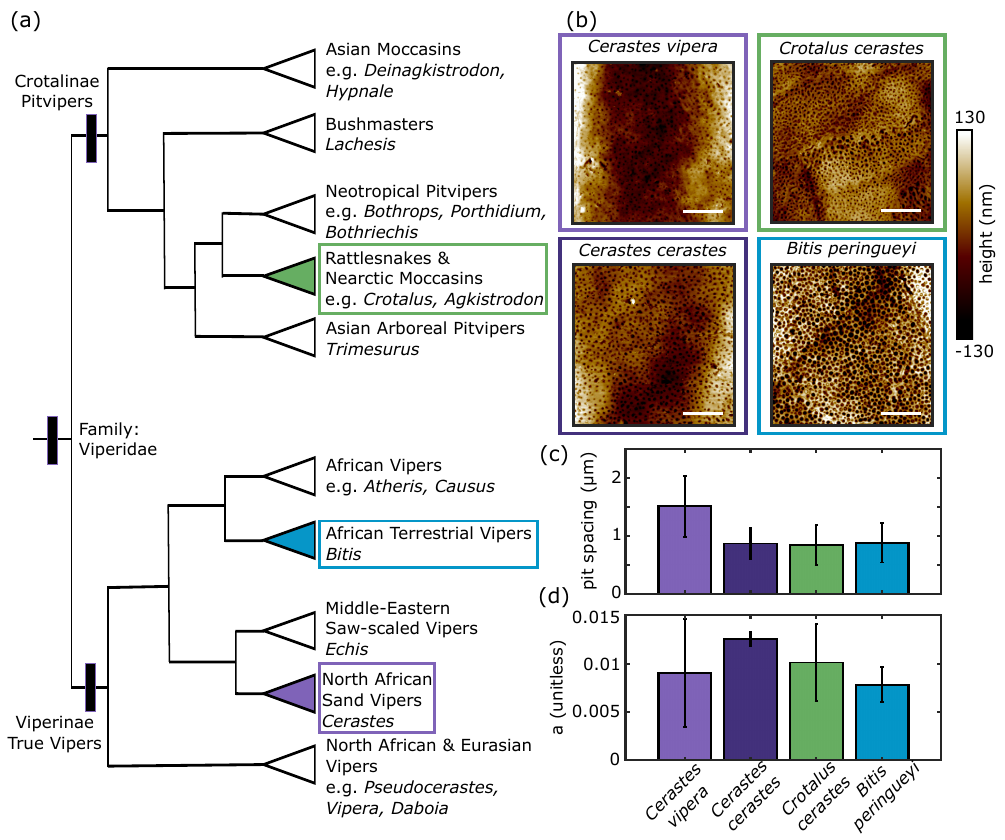}
    \caption{\textbf{A preserved \textit{B. peringueyi} sample reveals a third origin of convergently evolved microstructure in sidewinding vipers.} \textbf{(a)} A phylogeny of viperids adapted from~\cite{alencar_diversification_2016} shows the three branches represented in our sidewinder microstructure images. It is most parsimonious to conclude three independent origins of these microstructures, and sidewinding locomotion, than to propose a singular origin in the clade Viperidae with dozens of independent losses. \textbf{(b)} A similar pitted structure appears in each of the four sidewinding species imaged. Scale bars: 5~$\mu$m.\textbf{(c)} The spacing between pits in each species is similar, suggesting this structure may have evolved with a characteristic spacing. \textbf{(d)} The microstructure of each sidewinder species is similarly isotropic.}
    \label{fig:sidewinders}
\end{figure}

Using power spectra and Radon transforms (Figure~\ref{fig:SidewinderRawHeights}), we measured the characteristic spatial frequency of the pitted textures finding that \textit{Crotalus cerastes}, two species of \textit{Cerastes}, and our new results for \textit{B. peringueyi} all show an average pit spacing between $0.9$~$\mu$m and $1.5$~$\mu$m with overlapping variability up to $\pm0.6$~$\mu$m (Figure~\ref{fig:sidewinders}c). This overlap suggests convergence not only in structure type, but possibly also in pit spacing. Additionally, all sidewinding species show low anisotropy indices (Figure~\ref{fig:sidewinders}d), reflecting the isotropic nature of the pitted texture. These metrics are summarized in Table~\ref{table:allDataSidewinders}.Notably, \textit{B. peringueyi} conforms to the isotropic pitted pattern, providing another example of convergent evolution in both behavior and potentially beneficial ventral microstructures.

Previous work (Rieser et al., 2021) established that the microstructures associated with specialized sidewinding evolved independently in crotaline \textit{Crotalus cerastes} and the viperine genus \textit{Cerastes}. Our results here indicate a third evolutionary origin of the distinctive sidewinding microstructure in \textit{B. peringueyi} (Figure \ref{fig:sidewinders}a, which was adapted from~\cite{alencar_diversification_2016} and~\cite{tingle_morphological_2021}). This suggests the parsimonious hypothesis that similar microstructures have evolved once in the clade including \textit{Eristocophis macmahoni} and \textit{Pseudocerastes spp.} and again independently in \textit{Echis carinatus}. In other words, the five independent origins of specialized sidewinding in vipers~\cite{tingle_facultatively_2020} should be accompanied by five independent origins of sidewinding microstructures. Three origins of such microstructures have been confirmed (Rieser et al., 2021; this study), directing future work to investigate \textit{Echis}, \textit{Eristocophis}, and \textit{Pseudocerastes}.  
Collectively, these results suggest strong selective pressure for the performance-related microstructural trait among sidewinding vipers, likely in response to similar environmental challenges of sandy substrates~\cite{marvi_sidewinding_2014}. Nevertheless, sidewinding has evolved independently in desert-dwelling members of the Viperinae and Crotalinae without notable changes in other aspects of body morphology~\cite{tingle_facultatively_2020, tingle_morphological_2021}. Among the species studied here---three in Viperinae and one in Crotalinae---all exhibit similar pitted structures while micro-spikes are absent (Figure~\ref{fig:sidewinders}a,b). Anecdotal reports on non-specialized sidewinding exist for \textit{B. cornuta} and unclear reports of sidewinding---like behaviors are noted in \textit{B. arietans} and \textit{B. gabonica}. \textit{Bitis gabonica} is an exceptionally heavy bodied snake that is not associated with sandy substrates and is known to primarily use rectilinear locomotion (GWS, JRM, pers. obs.), so it is an unlikely candidate for either facultative sidewinding or microstructural adaptations for sidewinding. Yet, it does have notably reduced ventral micro-spikes (Rieser et al., 2021:fig. S1). Rieser et al. (2021:fig. S1) sampled three species of \textit{Echis} (but not the known sidewinding species \textit{E. carinatus}~\cite{tingle_facultatively_2020}) and found micro-spikes typical of non-sidewinding vipers in \textit{Echis leucogaster} and \textit{Echis coloratus} but an unusual, smooth texture in the non-sidewinding species \cite{tingle_facultatively_2020} \textit{Echis pyramidum}. Considered together, the results of this study along with Rieser et al., (2021) and Tingle (2020) indicate that more work on both sidewinding behaviors and ventral microstructures in viperines is warranted. Such work should consider morphology, behavior, and environment in an explicitly integrative evolutionary framework, as outlined by Tingle (2025)\cite{tingle_conceptual_2025}.

\section{Discussion and conclusions}

In this work, we used AFM to demonstrate that microstructures on the keratin-based skins of several preserved snakes and one lizard are each topographically similar to shed skin samples from the same species. Few studies have systematically quantified the in-plane geometry of snake microstructures; most have provided qualitative observations~\cite{arrigo_phylogenetic_2019,klein_ultrastructure_2014} or hand measured the size of features on the surface~\cite{rieser_functional_2021}. By collecting height profile data and calculating feature spacings, we developed robust analyses that enabled comparisons of ventral microstructures of reptiles from both shed skins and those found on preserved skins.

Our analyses revealed that preserved specimens, including those preserved for over $85$ years, retained their fine-scale surface structures. Specifically, we found that surface heights, spike spacings, and anisotropy indices were similar across sample types, with inter-sample variability falling within intra-sample variability. This validation greatly expands the range of specimens available for microstructural analysis beyond the limited supply and availability of shed skins and opens the door for broad studies linking structure to habitat, phylogeny, and locomotion strategy.

We leveraged our results to investigate an untested hypothesis~\cite{rieser_functional_2021} which discovered two evolutionarily independent losses of micro-spikes in favor of  isotropic pitted microstructures found on three sidewinding viper species. A computational model demonstrated that such isotropic structures may enhance frictional interactions that benefit sidewinding movement, and therefore, the authors predicted similar microstructures on a fourth sidewinding species, \textit{B. peringueyi}. While we were also unable to obtain a shed skin for this species, our results here indicate that preserved museum specimens can be used instead of shed skins. Indeed, AFM images of a museum specimen revealed that \textit{B. peringueyi} has the predicted pitted microstructure observed on the other sidewinding vipers. Comparisons of quantified structures demonstrated that all sidewinding species had a similar spacing between pits, suggesting that this structure may have evolved with a characteristic geometry.

We note that although the morphology of snake microstructures was retained in museum samples, 
the effects of preservation on mechanical properties---e.g., elasticity, friction, and wear---of the skin are not known. Though some prior work has measured elasticity~\cite{klein_epidermis_2012,klein_material_2010}, microscopic friction~\cite{benz2012anisotropic,baum_anisotropic_2014,hazel_nanoscale_1999,wu_variation_2020}, and microstructure wear and durability~\cite{klein_material_2010} for shed skins from a few species,  how structures relate to mechanical properties is not well understood. Future work should explore these connections further, as well as investigate how chemical fixatives used in preservation may or may not affect mechanical properties. Either way, linking microstructures to mechanics, even if only shed skin samples can be used, would not only advance our understanding of biological surface function but could also inspire the design of engineered materials with tunable anisotropic friction~\cite{wu_snakeinspired_2021,baum_dry_2014}.

While differences in mechanical properties across sample types remain to be explored, our results show that preserved specimens may be used in the place of shed skins to study structural properties of reptile skins. This finding greatly expands opportunities for high-resolution comparative studies across a wide range of species, including rare or elusive taxa. As the environment affects both the body shape~\cite{de_alencar_arboreality_2017} and skin thickness~\cite{shine_life_2019} of snakes, it is reasonable to hypothesize that the form and function of microstructures can vary as well. For example, dorsal microstructures in snakes often are hydrophobic and show strong phylogenetic signal~\cite{arrigo_phylogenetic_2019,spinner_non-contaminating_2014}. In contrast, ventral structures appear more closely tied to locomotion and ecological niche, such as observations of longer spikes on arboreal snakes (which may enhance frictional interactions) and shorter or smoother structures that may reduce drag or debris accumulation for aquatic and fossorial species~\cite{schmidt_snake_2012,gans_regional_1977,gower_scale_2003,martinez_quantifying_2021}. Museum collections can provide the phylogenetic diversity needed to gain new insights into drivers of the evolution of structures that are crucial for environmental interactions.

\section{Methods} \label{Methods}

\subsection{Sample preservation}
The preservation process involves injecting the animal with formalin (3.7\% formaldehyde), positioning it, and wrapping it in formalin-soaked cloth to cross-link proteins, position the specimen in a desired configuration, and prevent decomposition~\cite{dodd_preserving_2016,thavarajah2012chemical}. Samples are then preserved indefinitely in 70\% ethanol, which is relatively non-toxic and prevents microbial and fungal growth~\cite{simmons2024things}. Some specimens are preserved directly in ethanol, which better preserves DNA, though ethanol concentrations must remain high to prevent decay~\cite{dodd_preserving_2016}. Frozen samples were collected freshly deceased and kept on ice for two days, then submerged in water and placed in a freezer so ice would encase the specimen. Shed skins were collected from specimens housed at various locations, including the Rieser lab at Emory University and Zoo Atlanta in Atlanta GA (see Table~\ref{tab:sampleInfo}), and stored in sealed Ziplock bags at room temperature.

\subsection{Sample preparation}

For this study, we examined three different types of snake skin samples: those from shed skins, preserved samples, and frozen samples. For shed skins:  sample s2 (\textit{P. guttatus}) and the sample from \textit{O. attenuatus} were collected from our lab population after each animal had naturally shed; sample s1  (\textit{P. guttatus}) was collected at Zoo Atlanta, Georgia, along with samples from \textit{Crotalus polystictus} and \textit{Crotalus cerastes}; samples from \textit{Crotalus atrox} were obtained from a captive population in the Chiricahua Desert Museum in New Mexico. Sections of 2--3 ventral scales were cut out of the sheds for imaging.

Preserved samples were obtained from either the Field Museum of Natural History in Chicago, Illinois, or the Amphibian and Reptile Diversity Research Center, University of Texas at Arlington. These samples came from specimens that had been formalin fixed and preserved in $\geq$70\% ethanol. Samples from undamaged sections of freshly dead road-killed adult specimens of \textit{C. atrox} from Cochise County, Arizona, were formalin-fixed and ethanol-preserved using standard museum protocols; additional samples of skin from those specimens also were frozen for subsequent attempts at AFM imaging. To obtain skin from these samples for imaging, the top layer of skin on individual ventral scales was peeled off of the specimens. 

Initially, we attempted to image frozen and preserved samples while the skin was still attached to the scale tissue. This approach proved to be difficult because the sample would relax during imaging and the AFM tip would lose contact. This problem could potentially be remedied by letting the samples dry overnight or by imaging them while partially submerged in a bath of water (but with top surface exposed to air) but we found that these techniques worked inconsistently. The uppermost dermal layer, the stratum corneum, of skin would also occasionally become detached. As a result, we removed this layer and imaged it directly. Some museum samples that we obtained, however, did not have the top layer of skin present. We initially assumed that it was not removable on some samples, but after imaging the top layer of tissue on these samples, no microstructures were found. We concluded that the top layer of skin was absent in these cases; it is common to find the top layers of reptile scales detached from museum specimens and accumulated at the bottom of specimen jars. In order to accurately capture microstructures on museum or frozen samples, the stratum corneum layer of the epidermis must be present. The stratum corneum is the superficial skin layer and because it naturally comes loose in museum specimens, curators do not consider removal of the layer from a single scale to represent destructive sampling of specimens. Scale layers found already loose in the specimen jar may be used, but care must be taken to allocate a sample to a specific specimen if more than one specimen is housed in the jar.  

Before imaging, all skin samples were cleaned, using a previously established protocol~\cite{rieser_functional_2021}, to remove dirt and debris that may cover microstructures. The cut out or peeled off ventral scales were placed in a solution of 5\% Micro-90 mixed with deionized water. We then used a Cole Parmer ultrasonic cleaner to sonicate the samples in the solution for 30 minutes. After sonicating, the samples were rinsed with tap water to remove the Micro-90 solution. The samples were then laid flat on a wire screen and left to air dry in the lab for at least 12 hours to ensure they were completely dry before imaging. We used two-sided scotch tape to adhere the scales to glass microscope slides. After this, the samples were ready for imaging. We ensured that the cleaning and drying process does not affect the microstructure itself by imaging samples that were clean without washing. Nevertheless, small amounts of residual debris are expected and do appear in some of the analyzed images.

Our study does not capture variations in the structure based on the location on the body (reviewed in~\cite{schmidt_snake_2012},\cite{berthe_surface_2009}, and~\cite{klein_ultrastructure_2014}). One study found no variation in surface topography at different sites on the body in snakes~\cite{martinez_quantifying_2021}, comparing samples from different dorsal or ventral sites in individual specimens, but there are clear differences within individuals when comparing, for example in lizards, dorsal versus ventral surfaces as well as differences related to body size within species~\cite{baeckens_ontogenetic_2019}. Our goals were to compare microstructures in ventral scales among individuals of the same species using samples that were prepared or preserved differently and to compare microstructures on samples taken from adult snakes of comparable size within species; size-related differences in comparing different species are confounded by the fact that species naturally differ in size. Our study did not investigate any potential ontogenetic changes in species, nor any size-related variation in ventral microstructures that may occur within species. Further work that examines size and ontogenetic variation within species may prove interesting in the context of functionality.

Related specifically to museum specimens, detailed history of chemical methods of fixation and preservation often are not available for all specimens. Because fixation in approximately 10\% formalin solution and preservation in 70\% ethanol is traditional, we assume that our specimens borrowed from museums were prepared in this way. Our sampling and information available does not allow us to determine if microstructures persist under all possible methods of specimen preparation (e.g., fixation and preservation in isopropanol). All samples were taken from ventral scales in the middle third of each snake's body.

\subsection{Data collection: AFM imaging}

All samples were imaged using a Jupiter XR large-sample AFM from Asylum Research at Oxford Instruments. We used tapping mode imaging with either a 160AC or 240AC integrated silicon nitride probe. 20$\mu$m $\times$ 20$\mu$m and 60$\mu$m $\times$ 60$\mu$m sections of each sample were imaged with 512 pixel resolution at 1 Hz and 0.5 Hz speeds respectively. Snake skin is made of keratin~\cite{ripamonti_keratin-lipid_2009} and thus the imaging surfaces were very hard (elastic modulus of ~1-10 GPa)~\cite{mckittrick_structure_2012} and were undamaged by the imagining process, evidenced by taking 5-6 successive scans over the same region of the sample and observing no visual changes in the surface morphology.

To identify structural variation across the same sample, we imaged the same sample of shed skin from \textit{P. guttatus} at six locations across the same scale roughly spaced 5-mm apart laterally. All other images in this study were taken in a central region of a scale. To show the variation between different samples of from the same species, we imaged a second shed skin sample from \textit{P. guttatus}. We then compared structures across sample types by comparing these images to those taken from two different preserved samples of \textit{P. guttatus}. All images except those comparing variation across the same sample were taken in the central region of a scale. A similar series of images was taken for \textit{C. atrox} samples including frozen specimens and analyzed in Figure \ref{fig:atrox} to show the viability of frozen samples. 

\subsection{Analysis: Quantifying the structures} \label{Analysis: Quantifying the structures}

\begin{figure}
    \centering
    \includegraphics[width=\textwidth]{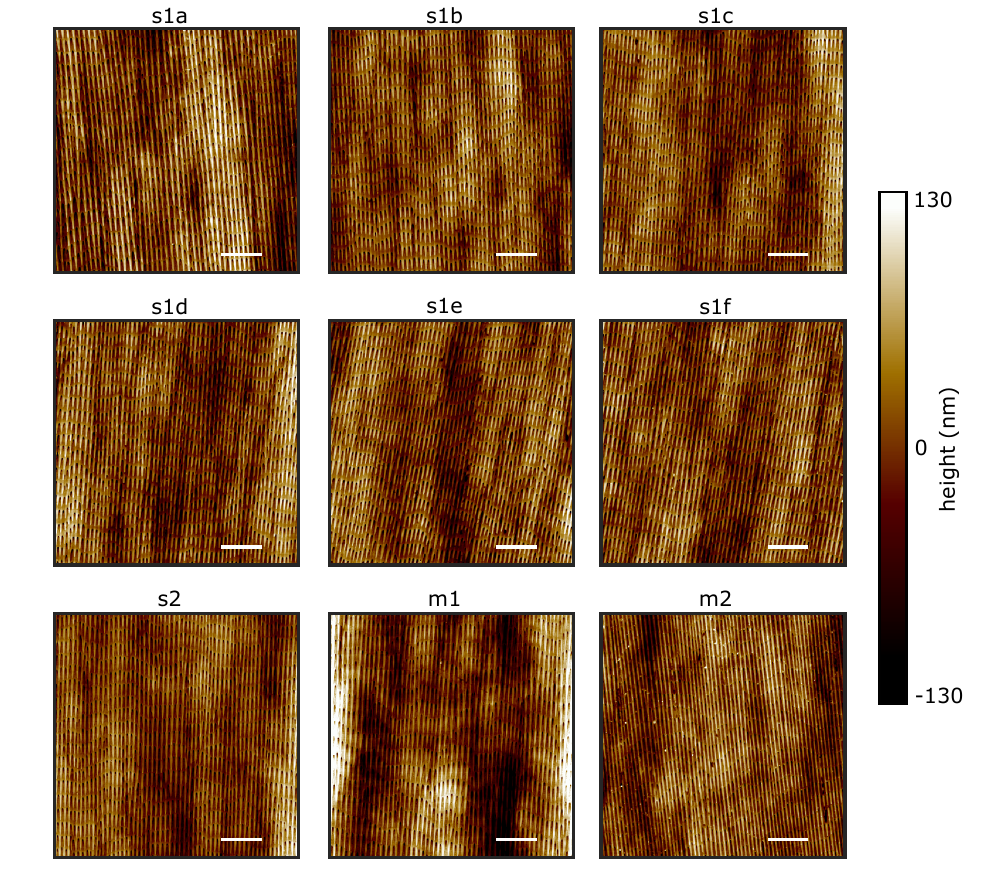}
    \caption{\textbf{Images from \textit{P. guttatus} samples before a Gaussian filter is applied.} Images s1a-s1f are from six sites across the same shed skin sample. s2 shows a second shed skin sample. m1 and m2 show museum samples from two individuals. Images are used for analysis in Figures \ref{fig:spacing} and \ref{fig:sample}. Scale bars: $10 \mu$m}
    \label{fig:OGims}
\end{figure}

\begin{figure}
    \centering
    \includegraphics[width=\textwidth]{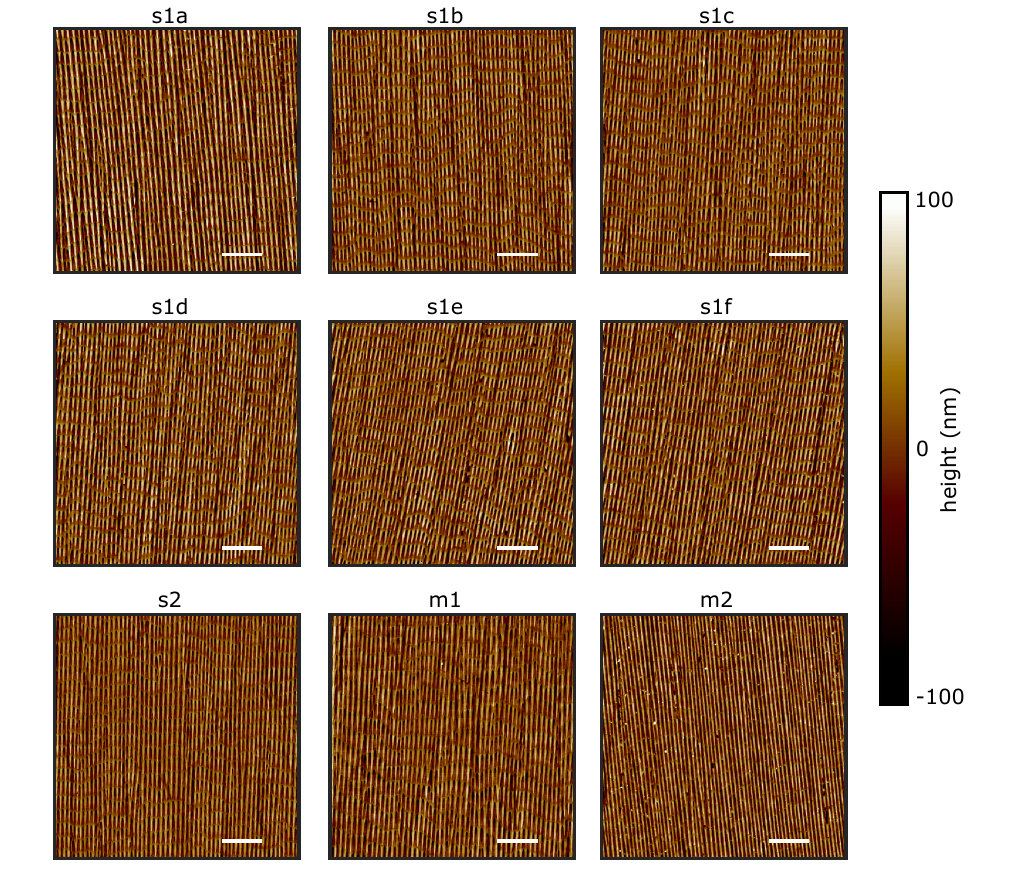}
    \caption{\textbf{Images from \textit{P. guttatus} samples after a Gaussian filter is applied.} Images correspond to the labeled images in Figure \ref{fig:OGims}. Images are used for analysis in Figures \ref{fig:spacing} and \ref{fig:sample}.  Scale bars: $10 \mu$m}
    \label{fig:ims}
\end{figure}

\begin{figure}
    \centering
    \includegraphics[width=\textwidth]{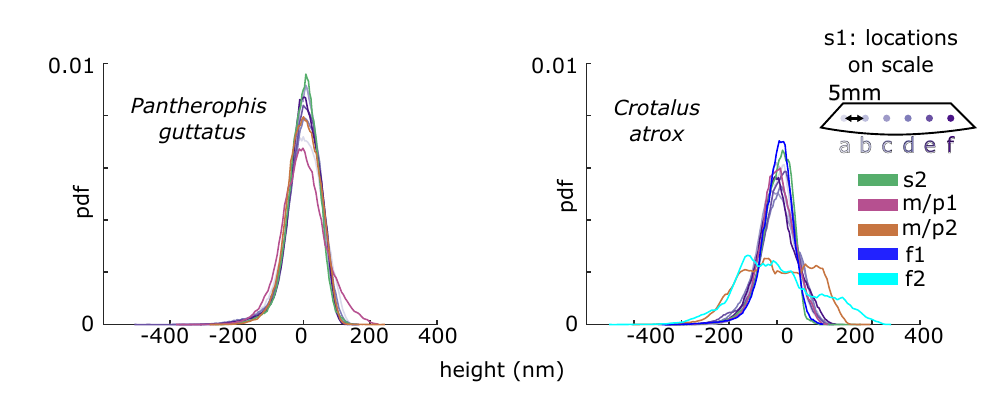}
    \caption{\textbf{Distributions of heights from unfiltered images.} Distributions from each image site in each sample from \textit{P. guttatus} and \textit{C. atrox} are shown.}
    \label{fig:RawHeights}
\end{figure}

Images from the AFM were flattened, meaning an offset was removed from each scan line so that they all have the same average value, and the overall slope of each scan line is removed. This algorithm is built in to the Igor-based AFM imaging software provided by Asylum Research. It is not a plane fit, instead acting on each scan line. The order of flattening can be selected where order one removes a linear fit, order two removes a quadratic, and so on. For each image, order one was chosen. Flattening accounts for thermal drift and sudden changes in tip behavior, both of which can introduce nonphysical offsets in height between consecutive scan lines and slopes in the sample. Flattening also corrects for small imperfections in sample mounting where it may be impossible to get the sample perfectly flat to the nanoscale. Flattened images were taken to be raw images for our purposes. The raw images still had larger scale depressions and bulges occurring over tens of microns, a larger scale than the spikes and pits we wanted to characterize. To bring the features of interest into the same plane, we applied a Gaussian filter with a standard deviation of 10 pixels to each image and subtracted the filter from the original image. The final step in processing was mean-subtracting each image. (See Figures~\ref{fig:OGims}, \ref{fig:ims}, \ref{fig:atroxImsOG}, and~\ref{fig:atroxIms} for the results of filtering images. Distributions of raw heights are shown in Figure~\ref{fig:RawHeights}.)

We first compared the distribution of heights in each of the filtered and mean subtracted images. By calculating the Jensen-Shannon (JS) divergence between pairs of distributions, we quantified how similar the heights of features and roughness are across images and sample types.

\begin{figure}
    \centering
    \includegraphics[width=\textwidth]{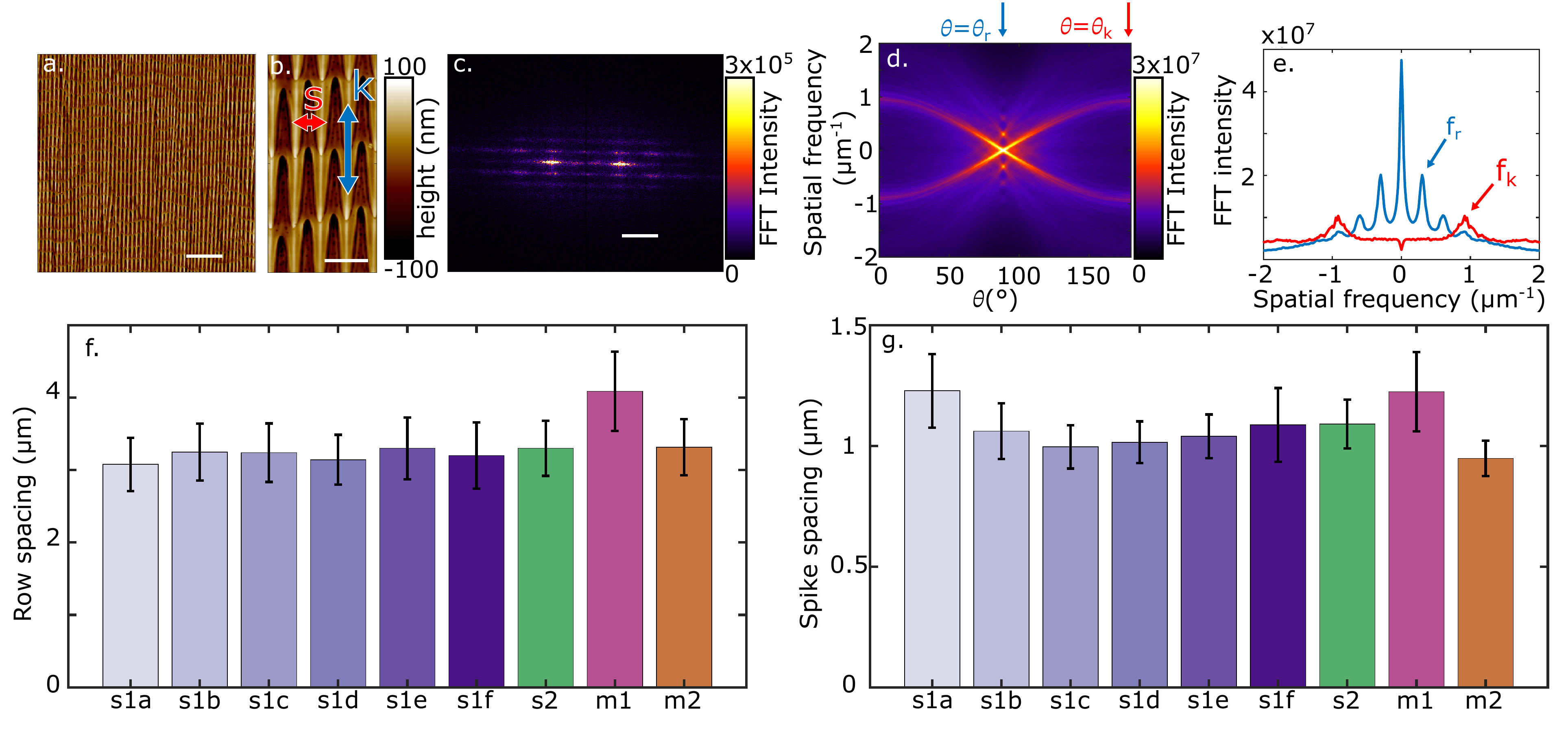}
    \caption{The same analysis in Figures \ref{fig:spacing} and \ref{fig:sample} but using a fast Fourier transform (FFT) rather than a power spectrum. \textbf{(a)} a 60x60 image of \textit{P. guttatus} microstrucutre. \textbf{(b)} a close-up showing the row spacing and spike spacing. \textbf{(c)} an FFT of the 60x60 image. \textbf{(d)} a Radon transform of the FFT and slices associated with each desired spacing. \textbf{(d)} the slices plotted showing the peaks used to find the row and spike spacing. \textbf{(e) and (f)} the row and spike spacing of each sample calculated similarly as when a power spectrum is used. Scale bars: \textbf{(a)} $10 \mu$m, \textbf{(b)} $2 \mu$m, \textbf{(c)} $1 \mu$m$^{-1}$,}
    \label{fig:wFFT}
\end{figure}

To calculate the periodicity of in-plane surface features, we built off a previous analysis technique~\cite{rieser_functional_2021}. We calculated a power spectrum of the height values in each image to identify dominant spatial frequencies. Using a fast Fourier transform rather than a power spectrum yielded similar results (Figure~\ref{fig:wFFT}). Images may be slightly tilted from their desired orientation, so we calculated a Radon transform with an angular increment of $0.25^\circ$ from these power spectra to more easily measure the location of dominant frequencies independent of the axes. A Radon transform projects the intensity of a $2$D image onto a single axis of varying orientation passing through the center of the image. For example, the Radon transform of a vertical line through the origin would have the line collapse to a single point summing all values along the line at $\theta=0^{\circ}$, then gradually increase in width until it reproduces the line at $\theta=90^{\circ}$ and then decreases in width back to a single point at $\theta=180^{\circ}$.

\begin{figure}
    \centering
    \includegraphics[height=20cm]{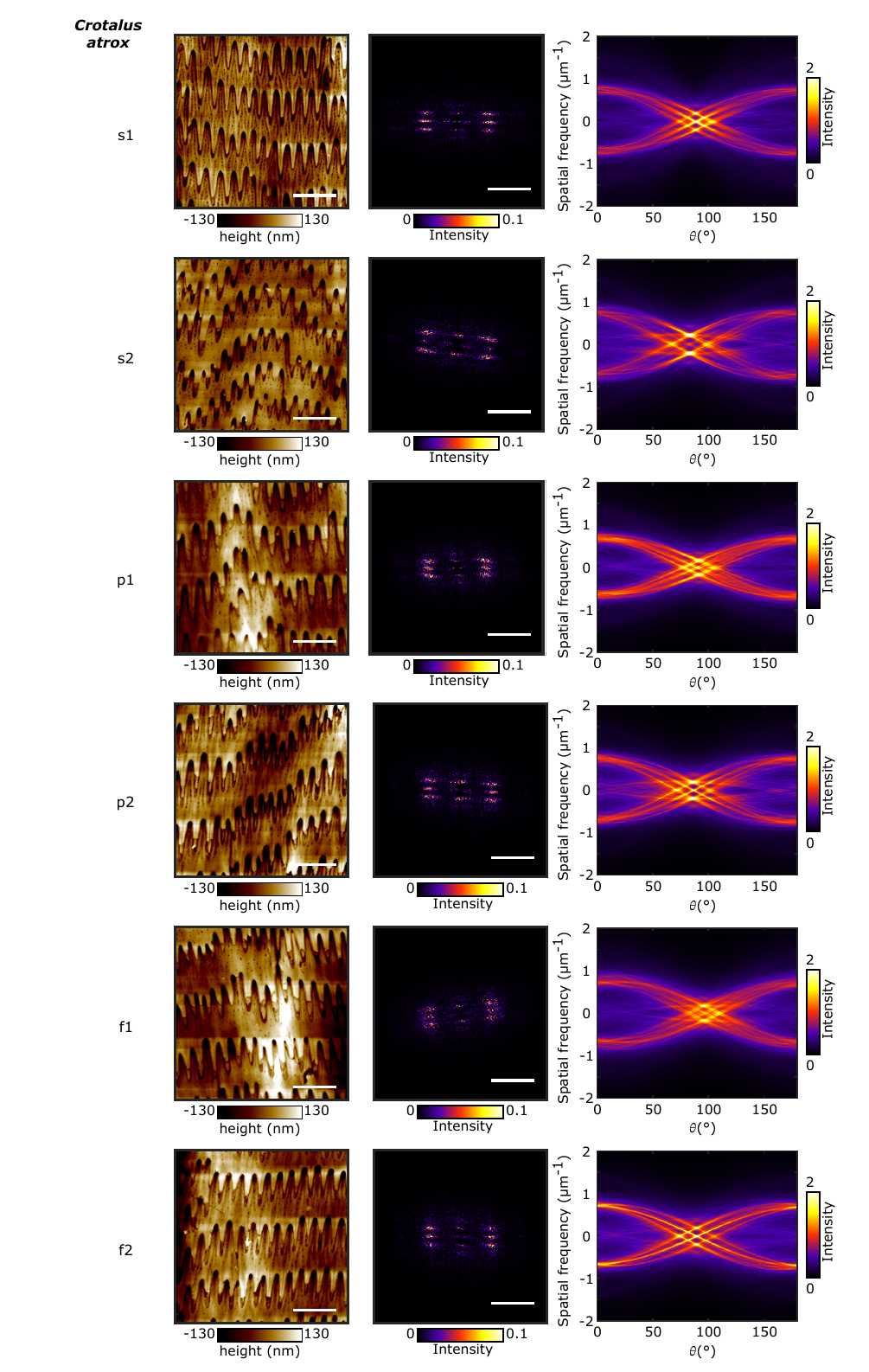}
    \caption{Images ($20\mu$m $\times 20\mu$m), power spectra, and Radon transforms from \textit{C. atrox} samples. Scale bars: AFM image (left column): $5\mu$m, power spectrum (middle column): $1\mu$m$^{-1}$}
    \label{fig:CroAtrRawHeights}
\end{figure}

\begin{figure}
    \centering
    \includegraphics[height=20cm]{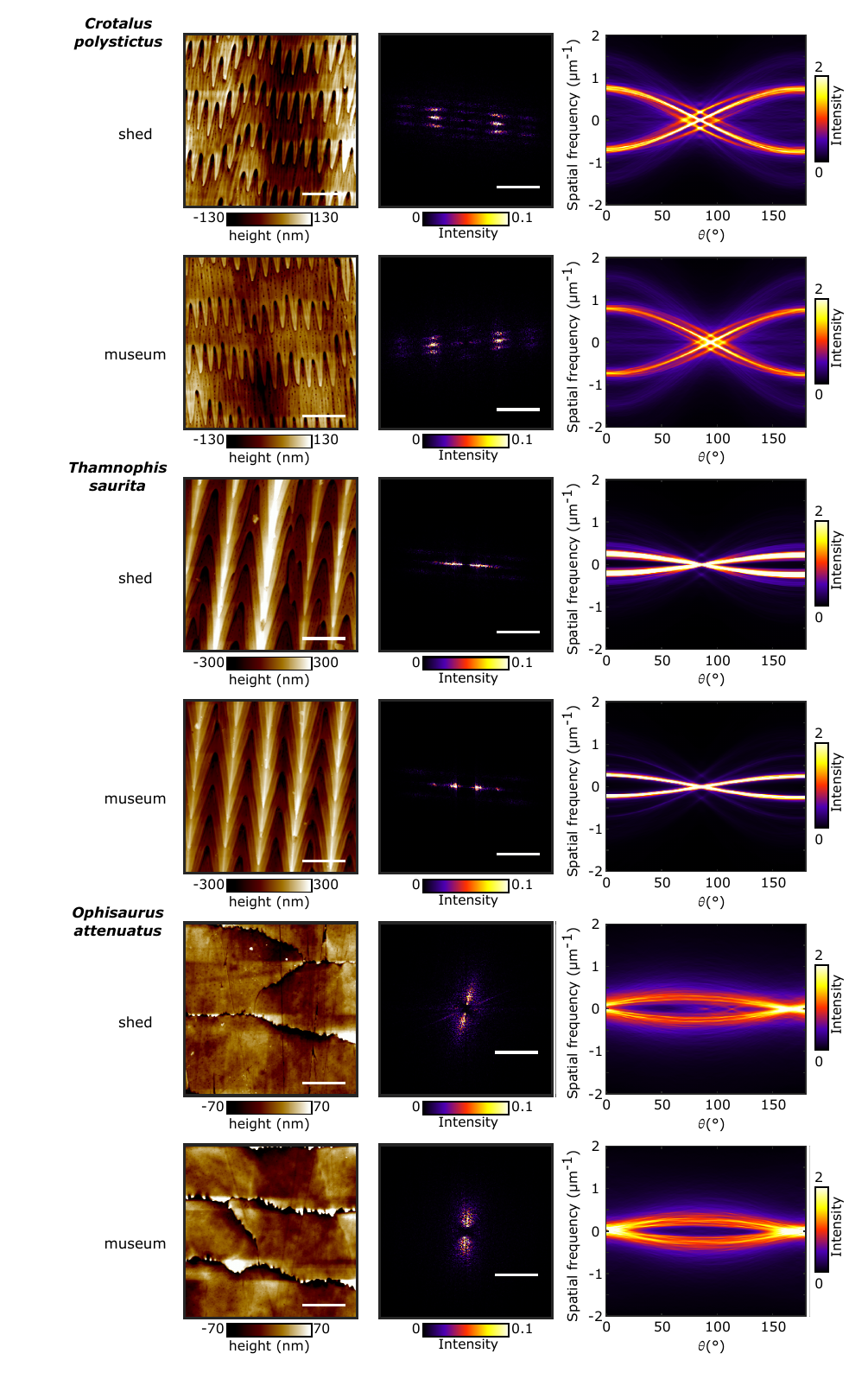}
    \caption{Images ($20\mu$m $\times 20\mu$m), power spectra, and Radon transforms from \textit{C. polystictus}, \textit{T. saurita}, and \textit{O. attenuatus} samples. Scale bars: AFM image (left column): $5\mu$m, power spectrum (middle column): $1\mu$m$^{-1}$}
    \label{fig:OtherRawHeights}
\end{figure}

\begin{figure}
    \centering
    \includegraphics[width=\textwidth]{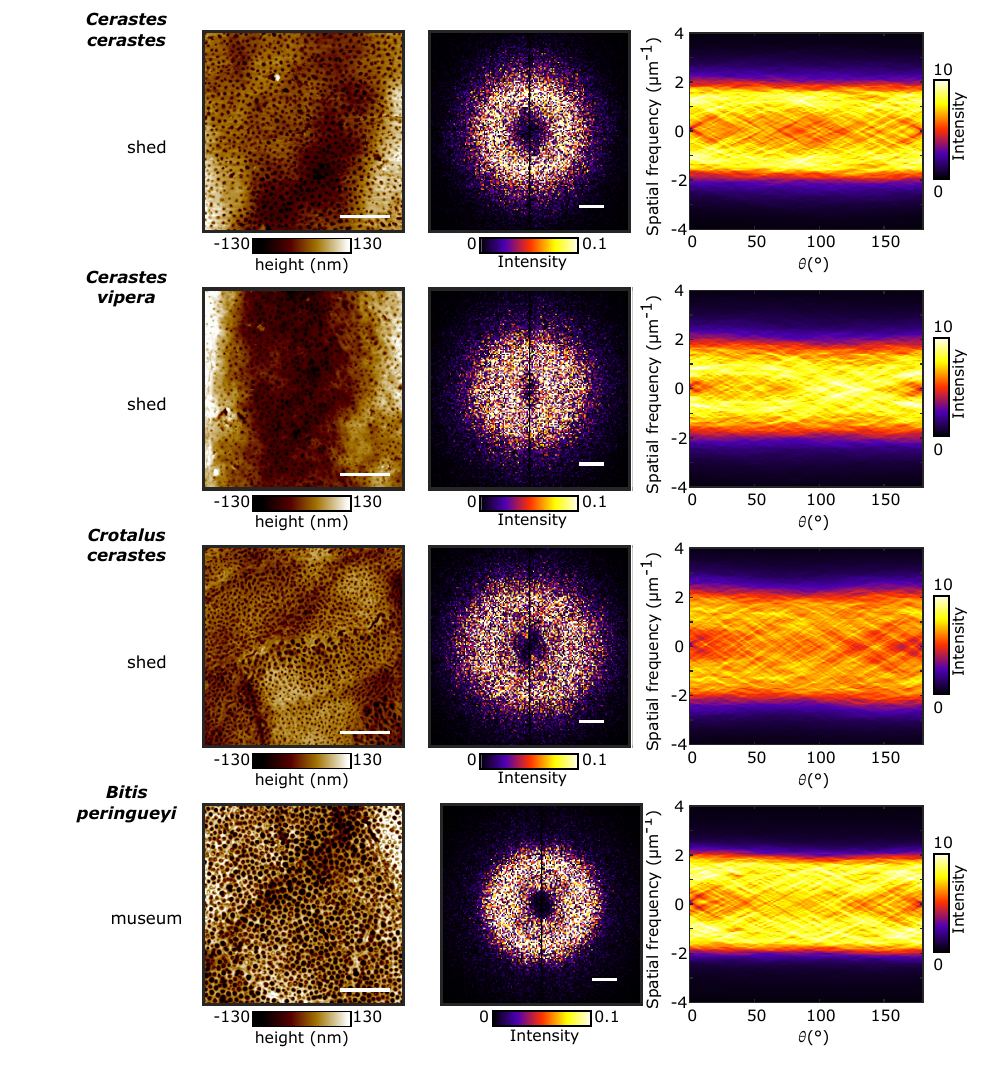}
    \caption{Images ($20\mu$m $\times 20\mu$m), power spectra, and Radon transforms from sidewinding species \textit{Cerastes cerastes}, \textit{Cerastes vipera}, \textit{Crotalus cerastes}, and \textit{Bitis peringueyi} samples. Scale bars: AFM image (left column): $5\mu$m, power spectrum (middle column): $1\mu$m$^{-1}$}
    \label{fig:SidewinderRawHeights}
\end{figure}

In order to accurately capture the spatial frequency of the structure, we imaged large regions that contained at least 15-20 rows of micro-spikes. For sidewinders, 20~$\mu$m~$\times$~20~$\mu$m images were deemed to have enough features to be adequate for this analysis. However, 20~$\mu$m~$\times$~20~$\mu$m images from non-sidewinding species only captured a few rows of spikes, so $60$~$\mu$m~$\times 60$~$\mu$~m images with several rows of spikes were used for this analysis. (Power spectra and Radon transforms for each of our samples are shown in Figures \ref{fig:PanGutRawHeights1}, \ref{fig:PanGutRawHeights2}, \ref{fig:CroAtrRawHeights}, \ref{fig:OtherRawHeights}, and \ref{fig:SidewinderRawHeights}.)

\begin{figure}
    \centering
    \includegraphics[height=11cm]{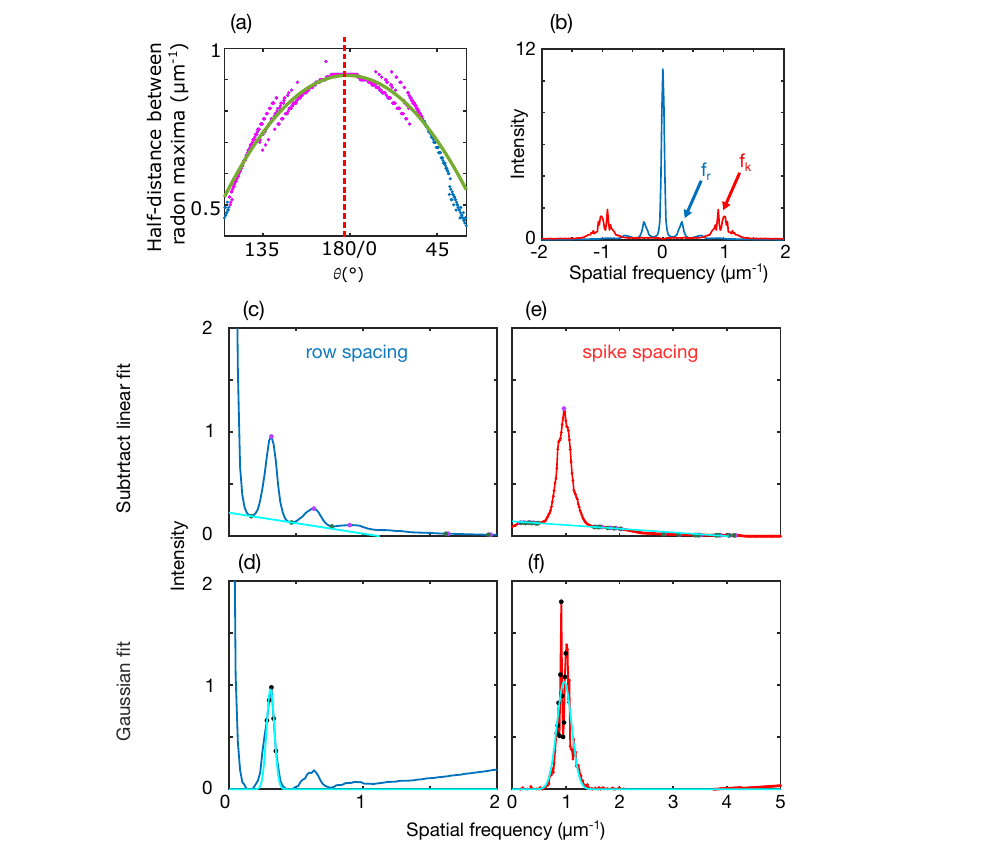}
    \caption{\textbf{Quantification of microscopic features shown in Figure\ref{fig:spacing}} \textbf{(a)} The distance between bands in each slice of the Radon transform are used to find the slice that gives the spatial frequency of spikes. At each angle value in the Radon transform, the half-distance between maxima on either side of zero was recorded and plotted. 400 points (magenta) around the maximum distance are fit with a parabola to account for noise. The angle value at the peak of the parabola is used to get the slice that gives the spike spacing. \textbf{(b)} Slices of the Radon transform used to calculate row spacing and spike spacing. The row spacing slice is the one that contains the maximum value in the whole Radon transform. \textbf{(c)} The peak of the Radon transform is set to a value of zero. To measure the row spacing, we take the positive half of the slice, smooth the data and find the first two local minima closest to zero. These minima surround a peak that gives the row frequently. We fit a linear model to these minima and subtract it from the data bringing the peak down to the x-axis to accurately assess its width. \textbf{(d)} We fit a Gaussian model to 5 points around the peak in the original, unsmoothed data. The peak position of the Gaussian is taken to be the row frequency and the standard deviation of the Gaussian is used as a measure of sample uncertainty. \textbf{(e)} A similar technique is applied to both sides of the spike spacing slice. Here a linear model is fit to the two minima surrounding the peak of the smoothed data and subtracted. \textbf{(f)} A Gaussian is fit to 11 points surrounding the peak in the unsmoothed data. After this method is repeated on the negative side of the slice, the spike frequency is taken to be the average absolute value of the two Gaussian peaks. The average standard deviation is used as a measure of sample uncertainty.}
    \label{fig:calculation}
\end{figure}

\begin{figure}
    \centering
    \includegraphics[width=\textwidth]{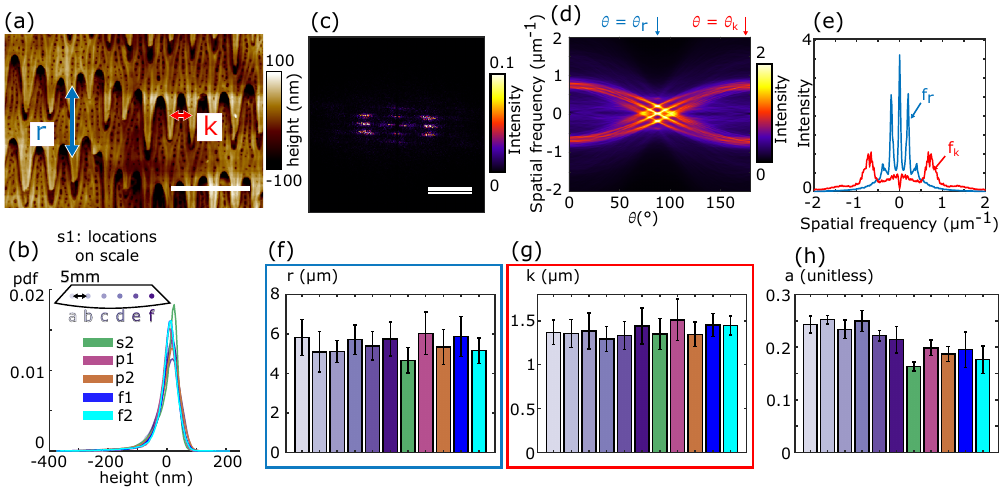}
    \caption{\textbf{The same analysis from Figures \ref{fig:spacing} and \ref{fig:sample} with samples from \textit{C. atrox}. This analysis also includes samples that are frozen and shows that microstructures are maintained on these as well.} \textbf{(a)} An image shows the details of the microstructure in sample s1. Height distributions from filtered $60\mu$m $\times 60\mu$m images at 6 sites in s1 as well as single sites in s2, p1, p2, f1, and f2 \textbf{(b)} are similar. A power spectrum \textbf{(c)} and Radon transform \textbf{(d)} are calculated from each image and the row spacing and spike spacing are calculated from their corresponding Radon transform slices according to the method outlined in Figure \ref{fig:calculation}. For this species, the row spacing \textbf{(f)} and spike spacing \textbf{(g)} are both retained in hand-preserved (p1, p2) as well as frozen (f1, f2) samples. The anisotropy \textbf{(h)}of each sample is slightly more variable but the anisotropy of the frozen and preserved samples are within the expected variation. Scale bars: \textbf{(a)} $2 \mu$m, \textbf{(b)} $1$~$\mu$m$^{-1}$}
    \label{fig:atrox}
\end{figure}

\begin{figure}
    \centering
    \includegraphics[width=\textwidth]{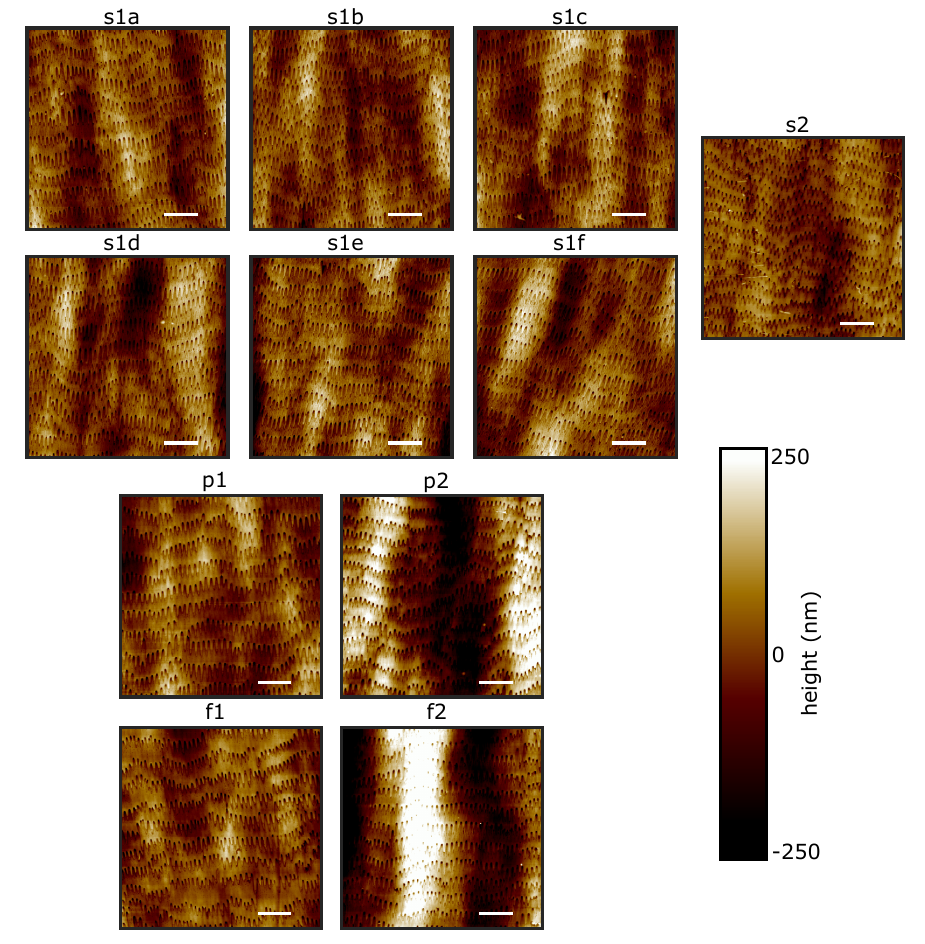}
    \caption{\textbf{Images from \textit{C. atrox} samples before a Gaussian filter is applied.} Images s1a-s1f are from six sites across the same shed skin sample. s2 shows a second shed skin sample. p1 and p2 show samples that were preserved in-house using the same technique museum samples from two individuals. Samples f1 and f2 were frozen after mortality and thawed before imaging. Images are used for analysis in Figure \ref{fig:atrox}. Scale bars: $10 \mu$m}
    \label{fig:atroxImsOG}
\end{figure}

\begin{figure}
    \centering
    \includegraphics[width=\textwidth]{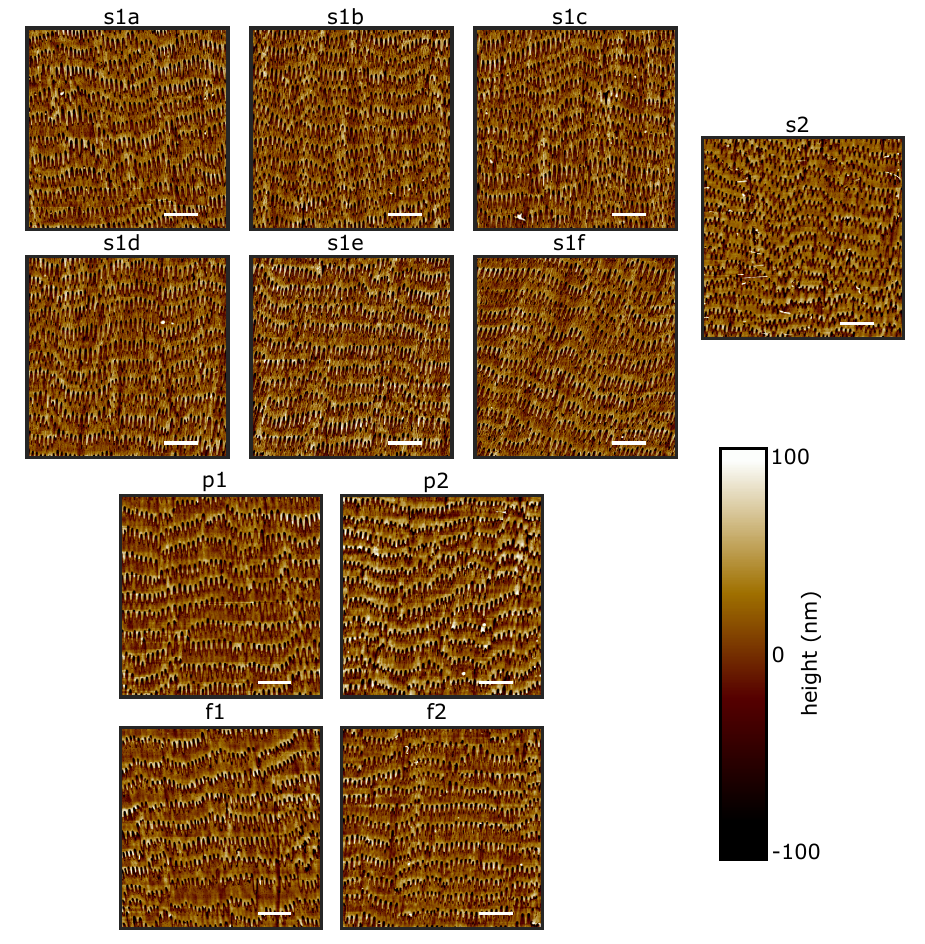}
    \caption{\textbf{Images from \textit{C. atrox} samples after a Gaussian filter is applied.} Images correspond to the labeled images in Figure \ref{fig:atroxImsOG}. Images are used for analysis in Figure \ref{fig:atrox}. Scale bars: $10 \mu$m}
    \label{fig:atroxIms}
\end{figure}

For spiked structures (all those on non-sidewinding snake species), the Radon transform slice with the highest peak has two secondary peaks whose frequency values correspond to the spacing between rows of spikes, $r$, in the image. The dominant peaks in the slice with two dominant peaks furthest apart corresponds to the lateral spacing between spikes, $k$ (see Figure~\ref{fig:calculation}a--b). In each case, the data was smoothed to identify the desired peaks and a linear model was fit to the minima on either side of the peak (Figure~\ref{fig:calculation}c,e). Subtracting the linear model from the slice allows us to fit a Gaussian curve to the peak and measure its width (Figure~\ref{fig:calculation}d,f). The mean of this distribution corresponds to the spatial frequency and the standard deviation gives a measure of the variation in the spatial frequency across the image. From these values we are able to calculate the row spacing, $r$, spike spacing $k$, and variability associated with each. 

Images from sidewinding species produced a much more uniform Radon transform. By taking an average of all slices in the Radon transform, we were able to identify mirrored peaks on either side of 0 Hz corresponding to a singular dominant spatial frequency. We subtracted the average Radon value at 0 Hz from the data and fit a Gaussian distribution to each side. We calculated the spacing between pits and the associated sample variability using the mean and standard deviation of these distributions.

We also adapted a previous technique~\cite{rieser_functional_2021} to calculate an anisotropy index, $a$, from each sample. We first divided each image into quadrants. We then calculated a power spectrum and Radon transform for each. From each Radon transform, we took a slice in the dominant $\theta_d$ direction containing the maximum intensity, and a second slice at $\theta_d\pm90^{\circ}$. Note that these slices are different from those used to calculate the row and spike spacing. Those slices tended to be $\approx90^\circ$ apart but were found independently. The JS divergence between the two slices serves as the anisotropy index for that quadrant. Finding the mean and standard deviation of the anisotropy indices gives a measure of the anisotropy index and its variability for the whole image.

\section*{Author contributions}C.A.R.: conceptualization, data curation, data interpretation, formal analysis, investigation, methodology, software, visualization, writing--original draft, writing--review and editing. G.W.S.: sample acquisition, writing--review and editing. J.R.M.III: conceptualization, resources, sample acquisition, writing--review and editing.  J.M.R.: conceptualization, data interpretation, methodology, project administration, resources, software, supervision, writing--review and editing.



\section*{Acknowledgments}The authors thank the Field Museum of Natural History The University of Texas at Arlington for loans of specimens; Jessica Tingle and Gordon Berman for helpful discussions.

\pagebreak

\beginsupplement
\newpage
\setcounter{page}{1}
\section{Supplementary Material}

\begin{table}[ht]
    \centering
    \tiny
    \caption{Sample information. *Preserved in house by Joseph R. Mendelson III}
    \begin{tabular}{c||c|c|c}
        \textbf{Species} & \textbf{Sample type} & \textbf{Year collected} & \textbf{Source} \\
        \hline
        \hline
        \textit{Pantherophis guttatus} & shed skin (s1) & 2021 & Zoo Atlanta, Atlanta GA\\
        \hline
        \textit{Pantherophis guttatus} & shed skin (s2) & 2024 & Rieser Lab, Emory University, Atlanta GA\\
        \hline
        \textit{Pantherophis guttatus} & museum (m1) & unknown & University of Texas at Arlington, Arlington TX\\
        \hline
        \textit{Pantherophis guttatus} & museum (m2) & 1964 & University of Texas at Arlington, Arlington TX\\
        \hline
        \textit{Crotalus polystictus} & shed skin & 2024 & Zoo Atlanta, Atlanta GA\\
        \hline
        \textit{Crotalus polystictus} & museum & 1939 & Field Museum of Natural History, Chicago IL\\
        \hline
        \textit{Crotalus atrox} & shed skin (s1) & unknown & Courtesy of Gordon Schuett\\
        \hline
        \textit{Crotalus atrox} & shed skin (s2) & unknown & Courtesy of Gordon Schuett\\
        \hline
        \textit{Crotalus atrox} & museum* (p1) & 2021 & Courtesy of Gordon Schuett\\
        \hline
        \textit{Crotalus atrox} & museum* (p2) & 2021 & Courtesy of Gordon Schuett\\
        \hline
        \textit{Crotalus atrox} & frozen (f1) & 2021 & Courtesy of Gordon Schuett\\
        \hline
        \textit{Crotalus atrox} & frozen (f2) & 2021 & Courtesy of Gordon Schuett\\
        \hline
        \textit{Thamnophis saurita} & shed skin & unknown & Joseph R. Mendelson III\\
        \hline
        \textit{Thamnophis saurita} & museum & 1971 & University of Texas at Arlington, Arlington TX\\
        \hline
        \textit{Ophisaurus attenuatus} & shed skin & 2022 & Rieser Lab, Emory University, Atlanta GA\\
        \hline
        \textit{Ophisaurus attenuatus} & museum & 1994 & University of Texas at Arlington, Arlington TX\\
        \hline
        \textit{Cerastes cerastes} & shed skin & unknown & Joseph R. Mendelson III\\
        \hline
        \textit{Cerastes vipera} & shed skin & unknown & Joseph R. Mendelson III\\
        \hline
        \textit{Crotalus cerastes} & shed skin & unknown & Zoo Atlanta, Atlanta GA\\
        \hline
        \textit{Bitis peringueyi} & museum & 1987 & University of Texas at Arlington, Arlington TX\\
    \end{tabular}
    \label{tab:sampleInfo}
\end{table}

\begin{table}[ht]
\centering
\caption{JS divergence values between each pair of height distributions analyzed for each \textit{P. guttatus} image. Note that JS divergence values between museum samples and shed skin samples are low and comparable to the values between shed skins indicating a high degree of similarity.}
\begin{tabular}{c|c|c|c|c|c|c|c|c|c}
samples & \textbf{s1a} & \textbf{s1b} & \textbf{s1c} & \textbf{s1d} & \textbf{s1e} & \textbf{s1f} & \textbf{s2} & \textbf{m1} & \textbf{m2} \\
\hline
\textbf{s1a} &0 & 0.0273 & 0.0393 & 0.0157 & 0.0185 & 0.0104 & 0.0423 & 0.0332 & 0.0149 \\
\hline
\textbf{s1b} &0.0273 & 0 & 0.0021 & 0.0048 & 0.0018 & 0.0054 & 0.0076 & 0.0042 & 0.0184 \\
\hline
\textbf{s1c} &0.0393 & 0.0021 & 0 & 0.0092 & 0.0054 & 0.0122 & 0.0069 & 0.0051 & 0.0277 \\
\hline
\textbf{s1d} &0.0157 & 0.0048 & 0.0092 & 0 & 0.0012 & 0.0029 & 0.0193 & 0.0124 & 0.0186 \\
\hline
\textbf{s1e} &0.0185 & 0.0018 & 0.0054 & 0.0012 & 0 & 0.0022 & 0.0125 & 0.0074 & 0.0155 \\
\hline
\textbf{s1f} &0.0104 & 0.0054 & 0.0122 & 0.0029 & 0.0022 & 0 & 0.0162 & 0.0114 & 0.0095 \\
\hline
\textbf{s2} &0.0423 & 0.0076 & 0.0069 & 0.0193 & 0.0125 & 0.0162 & 0 & 0.0039 & 0.0203 \\
\hline
\textbf{m1} &0.0332 & 0.0042 & 0.0051 & 0.0124 & 0.0074 & 0.0114 & 0.0039 & 0 & 0.0175 \\
\hline
\textbf{m2} &0.0149 & 0.0184 & 0.0277 & 0.0186 & 0.0155 & 0.0095 & 0.0203 & 0.0175 & 0 \\
\end{tabular}
\label{table:heightJS}
\end{table}


\begin{figure}
    \centering
    \includegraphics[height=20cm]{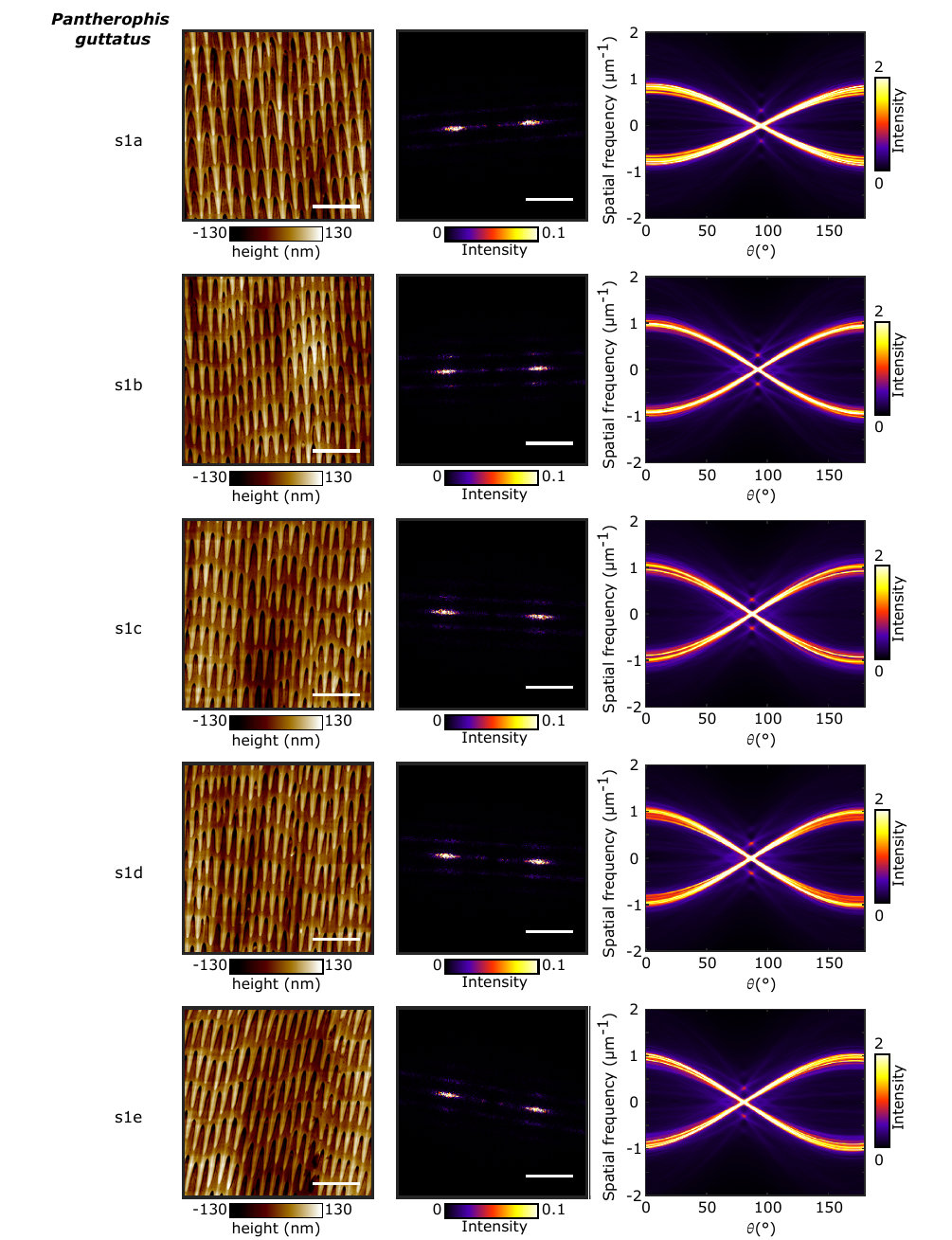}
    \caption{Images ($20\mu$m $\times 20\mu$m), power spectra, and Radon transforms from \textit{P. guttatus} samples. Scale bars: AFM image (left column): $5\mu$m, power spectrum (middle column): $1\mu$m$^{-1}$}
    \label{fig:PanGutRawHeights1}
\end{figure}

\begin{figure}
    \centering
    \includegraphics[width=\textwidth]{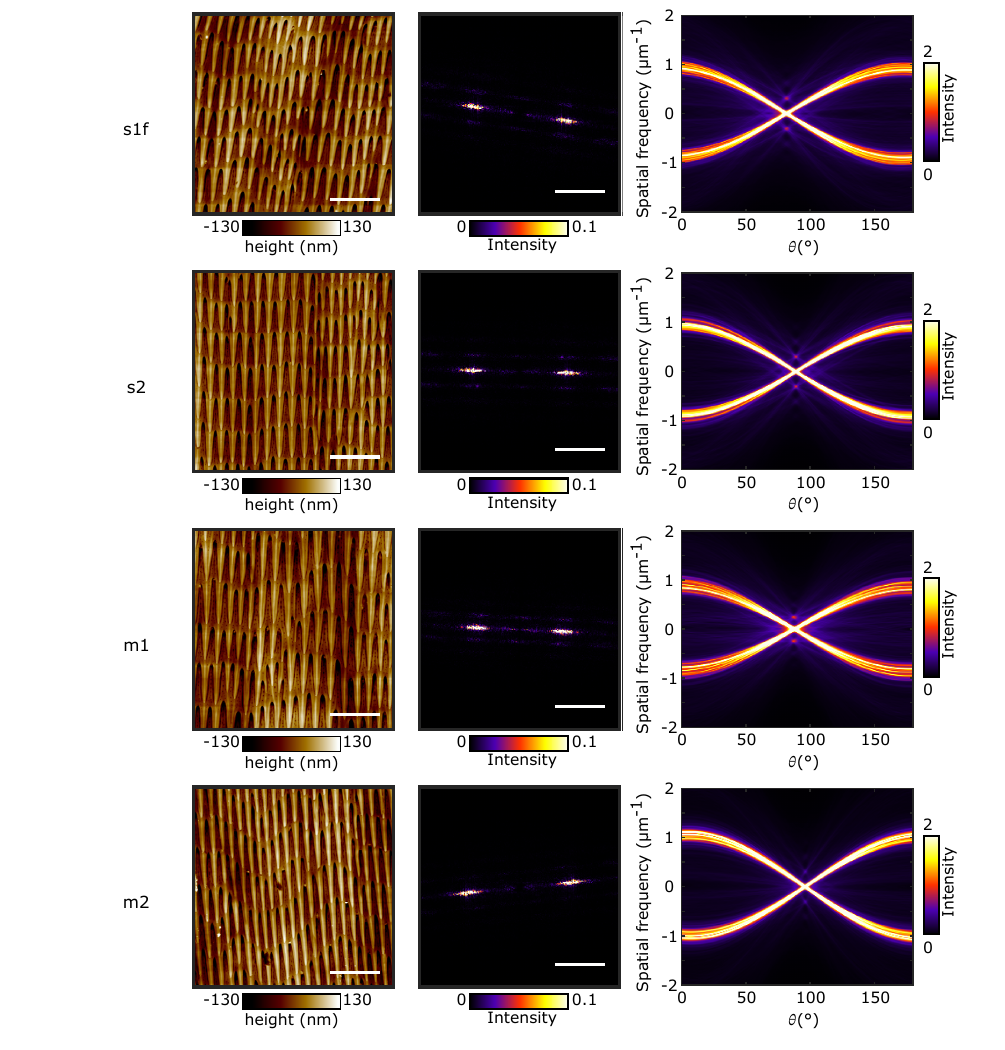}
    \caption{Cont. from Figure \ref{fig:PanGutRawHeights1}: Images ($20\mu$m $\times 20\mu$m), power spectra, and Radon transforms from \textit{P. guttatus} samples. Scale bars: AFM image (left column): $5\mu$m, power spectrum (middle column): $1\mu$m$^{-1}$}
    \label{fig:PanGutRawHeights2}
\end{figure}

\begin{table}
\centering
\tiny
\caption{Spacing data for species where multiple sample types were studied. Most have spiked microstructures and strong periodicity in two directions. \textit{O. attenuatus} only had strong periodicity in one direction corresponding to the rows of cells.}
\begin{tabular}{c|c|c|c|c|c|c|c|c|c}
species & sample/site label & row spacing $r$ & $\sigma_r$ & $\sigma_{rfit}$ & spike spacing $k$ & $\sigma_k$ & $\sigma_{kfit}$ & anisotropy & $\sigma_a$\\
\hline
\textit{P. guttatus} & s1a & 3.076 & 0.284 & 0.055 & 1.224 & 0.089 & 0.025 & 0.587 & 0.037 \\
\hline
\textit{P. guttatus} & s1b & 3.249 & 0.303 & 0.045 & 1.063 & 0.058 & 0.011 & 0.535 & 0.019 \\
\hline
\textit{P. guttatus} & s1c & 3.227 & 0.305 & 0.036 & 1.032 & 0.141 & 0.136 & 0.507 & 0.014 \\
\hline
\textit{P. guttatus} & s1d & 3.140 & 0.288 & 0.049 & 1.006 & 0.043 & 0.004 & 0.540 & 0.016 \\
\hline
\textit{P. guttatus} & s1e & 3.295 & 0.353 & 0.037 & 1.011 & 0.058 & 0.009 & 0.514 & 0.012 \\
\hline
\textit{P. guttatus} & s1f & 3.222 & 0.406 & 0.054 & 1.111 & 0.084 & 0.013 & 0.542 & 0.011 \\
\hline
\textit{P. guttatus} & s2 & 3.288 & 0.293 & 0.022 & 1.099 & 0.070 & 0.010 & 0.621 & 0.024 \\
\hline
\textit{P. guttatus} & m1 & 4.101 & 0.443 & 0.038 & 1.228 & 0.110 & 0.033 & 0.595 & 0.014 \\
\hline
\textit{P. guttatus} & m2 & 3.304 & 0.324 & 0.017 & 0.957 & 0.083 & 0.022 & 0.685 & 0.030\\
\hline
\textit{C. atrox} & s1a & 5.821 & 0.916 & 0.017 & 1.370 & 0.141 & 0.024 & 0.243 & 0.016 \\
\hline
\textit{C. atrox} & s1b & 5.097 & 1.029 & 0.318 & 1.354 & 0.161 & 0.011 & 0.252 & 0.008 \\
\hline
\textit{C. atrox} & s1c & 5.115 & 0.567 & 0.043 & 1.385 & 0.204 & 0.082 & 0.234 & 0.018 \\
\hline
\textit{C. atrox} & s1d & 5.712 & 0.749 & 0.181 & 1.293 & 0.140 & 0.009 & 0.250 & 0.020 \\
\hline
\textit{C. atrox} & s1e & 5.403 & 0.728 & 0.040 & 1.331 & 0.160 & 0.023 & 0.222 & 0.010 \\
\hline
\textit{C. atrox} & s1f & 5.746 & 0.867 & 0.167 & 1.441 & 0.204 & 0.018 & 0.214 & 0.025 \\
\hline
\textit{C. atrox} & s2 & 4.664 & 0.640 & 0.102 & 1.350 & 0.175 & 0.043 & 0.163 & 0.009 \\
\hline
\textit{C. atrox} & p1 & 6.040 & 1.073 & 0.446 & 1.511 & 0.237 & 0.023 & 0.199 & 0.015 \\
\hline
\textit{C. atrox} & p2 & 5.348 & 0.879 & 0.242 & 1.348 & 0.137 & 0.005 & 0.187 & 0.014 \\
\hline
\textit{C. atrox} & f1 & 5.865 & 1.011 & 0.312 & 1.453 & 0.134 & 0.013 & 0.196 & 0.034 \\
\hline
\textit{C. atrox} & f2 & 5.158 & 0.648 & 0.111 & 1.447 & 0.109 & 0.017 & 0.176 & 0.026 \\
\hline
\textit{C. polystictus} & shed & 5.225 & 0.683 & 0.070 & 1.376 & 0.129 & 0.018 & 0.405 & 0.019 \\
\hline
\textit{C. polystictus} & museum & 6.106 & 0.865 & 0.064 & 1.319 & 0.103 & 0.008 & 0.336 & 0.006 \\
\hline
\textit{T. saurita} & shed & 4.273 & 0.542 & 0.035 & 4.304 & 0.743 & 0.078 & 0.614 & 0.024 \\
\hline
\textit{T. saurita} & museum & 3.749 & 0.465 & 0.045 & 4.008 & 0.167 & 0.018 & 0.705 & 0.012 \\
\hline
\textit{O. attenuatus} & shed & 3.640 & 1.269 & 0.098 & n/a & n/a & n/a & 0.103 & 0.040 \\
\hline
\textit{O.attenuatus} & museum & 3.773 & 1.751 & 0.143 & n/a & n/a & n/a & 0.152 & 0.011 \\
\end{tabular}
\label{table:allData}
\end{table}



\begin{table}
\centering
\caption{Spacing data for sidewinder samples.}
\begin{tabular}{c|c|c|c|c|c|c}
species & sample type & pit spacing & $\sigma$ & $\sigma_{fit}$ & anisotropy & $\sigma$\\
\hline
\textit{Cerastes cerastes} & shed & 0.870 & 0.272 & 0.009 & 0.013 & 0.001 \\
\hline
\textit{Cerastes vipera} & shed & 1.509 & 0.530 & 0.022 & 0.009 & 0.006 \\
\hline
\textit{Crotalus cerastes} & shed & 0.843 & 0.339 & 0.009 & 0.010 & 0.004 \\
\hline
\textit{Bitis peringueyi} & museum & 0.880 & 0.338 & 0.016 & 0.008 & 0.002 \\
\end{tabular}
\label{table:allDataSidewinders}
\end{table}

\end{document}